\documentclass[a4paper,aps,prb,reprint,showpacs,twocolumns]{revtex4-1}
\usepackage{amsmath,amssymb,mathrsfs}
\usepackage[pdftex]{graphicx}
\usepackage{color}

\newcommand{\ket}[1]{\left|#1\right\rangle}
\newcommand{\bra}[1]{\left\langle#1\right|}

\newcommand{\ii}{\text{i}}
\newcommand{\dd}{\text{d}}

\newcommand{\comm}[2]{\left[#1,#2\right]}

\allowdisplaybreaks[4]

\begin{document}

\bibliographystyle{unsrt}

\title{Time evolution during and after finite-time quantum quenches in Luttinger liquids}
\author{Piotr Chudzinski}
\affiliation{Institute for Theoretical Physics, Center for Extreme Matter and Emergent Phenomena, Utrecht University, Princetonplein 5, 3584 CE Utrecht, The Netherlands}
\author{Dirk Schuricht}
\affiliation{Institute for Theoretical Physics, Center for Extreme Matter and Emergent Phenomena, Utrecht University, Princetonplein 5, 3584 CE Utrecht, The Netherlands}
\date{August 10th 2016}
\pagestyle{plain}

\begin{abstract}
We consider finite-time quantum quenches in the interacting Tomonaga--Luttinger model, for example time-dependent changes of the nearest-neighbour interactions for spinless fermions. We use the exact solutions for specific protocols including the linear and cosine ramps (or, more generally, periodic pumping). We study the dynamics of the total and kinetic energy as well as the Green functions during as well as after the quench. For the latter we find that the light-cone picture remains applicable, however, the propagating front is delayed as compared to the sudden quench. We extract the universal behaviour of the Green functions and in particular provide analytic, non-perturbative results for the delay applicable to quenches of short to moderate duration but arbitrary time dependency.
\end{abstract}
\maketitle

\section{Introduction}\label{sec:intro}
Over the last decade, due to advances\cite{Bloch-08,ReichelVuletic11,Langen-15} in the control of ultra-cold atomic systems the experimental realisation of sudden quantum quenches\cite{CalabreseCardy06} has become possible. Hereby a quantum system is prepared in an initial state, say the ground state of an Hamiltonian $H_0$, and then its time evolution after switching to another Hamiltonian $H$, eg, obtained by suddenly switching one of the system parameters of $H_0$, is studied. Generically the initial state will have a highly non-trivial representation in terms of the eigenstates of $H$, thus resulting in a complicated relaxation of observables after the quench. These experimental advances have triggered tremendous theoretical efforts\cite{Polkovnikov-11,Eisert-15,GogolinEisert16} to understand the quench dynamics of a vast class of systems covering all spatial dimensions.

Of particular interest have been one-dimensional systems, due to the availability of powerful numerical and analytical tools as well as their relation to topics like integrability. Arguably the most generic one-dimensional system is the Luttinger liquid\cite{Giamarchi04,Cazalilla-11,Schoenhammer13} which is known to describe the low-energy properties of gapless systems like quantum wires, spin chains or bosonic atoms in one-dimensional optical lattices. The importance of Luttinger liquids motivated the detailed investigation\cite{Cazalilla06,Perfetto06,DeChiara-06,Barmettler-09,Uhrig09,IucciCazalilla09,Barmettler-10,MitraGiamarchi11,Mitra12,KRSM12,MitraGiamarchi12,RSM12,DallaTorre-13,Mitra13,NessiIucci13,KennesMeden13,NgoDinh-13,HamerlaUhrig13,Coira-13,Tavora-14,Kennes-14,Protopopov-15,Collura-15} of its relaxation dynamics after sudden quantum quenches.  

While the vast majority of previous works focused on sudden quantum quenches, we will here consider more general quenches of finite length $\tau$ over which the system parameters vary (see Fig.~\ref{fig:nonsudden} for a sketch). Physically the finite quench time $\tau$ (depending on the precise form of the quench protocol one may even introduce several time scales $\tau_n$) introduces an additional energy scale $\Delta_\text{quench}\sim 1/\tau$, which is obviously trivial in the sudden and adiabatic limit. This newly generated energy scale is directly related to the quench protocol, ie, the switching process, and thus can be tuned. In particular, it can be made comparable to the other energy scales in the system like the band width, excitations gaps or relaxation rates. The interplay of these different energy scales originating in the properties of the post-quench Hamiltonian, the initial state and the quench protocol opens the possibility to study emergent quantum states beyond the ones accessible via sudden quench protocols. An example of such an emergent state is generated by the periodic quench discussed below.

Several aspects of finite-time quenches and the interpolation between the sudden and adiabatic limit have been studied.\cite{PolkovnikovGritsev08,EcksteinKollar10,MoeckelKehrein10,Bernier-11,TomarasKehrein11,Bernier-12,Sandri-12,HaqueZimmer13,Das-15} For the Luttinger liquid Dora et al.\cite{Dora-11} first considered linear quench protocols. They obtained perturbative results for the total energy and fermionic chiral Green function, which were later extended to spin-spin correlation functions and compared with numerical simulations of the time evolution during the quench in the XXZ Heisenberg chain.\cite{Pollmann-13} Bernier et al.\cite{Bernier-14} went beyond the perturbative treatment by deriving exact results for the time evolution during linear quenches in terms of Bessel functions (see Sec.~\ref{sec:linear}). This enabled them to analyse the properties of the bosonic Green function in great detail, including the derivation of power laws governing the propagation of the light cone during the quench. The obtained results were further confirmed by numerical simulations for the Bose--Hubbard model. Further aspects that were investigated include the excitation energy, the work statistics, finite-temperature initial states, the Loschmidt echo and the diagonal ensemble reached at late times.\cite{DziarmagaTylutki11,PerfettoStefanucci11,Dora-12,Dora-13,BacsiDora13,Sachdeva-14,Porta-16}

In this article we aim at obtaining a complete understanding of finite-time quenches in Luttinger liquids. To this end we consider the time evolution during and after the quench and derive exact, analytical results to go beyond the perturbative regime. This allows us to study the interplay between the quench time $\tau$ and other energy scales in the system, resulting for example in a non-trivial dependence of the total energy on the quench time (see Fig.~\ref{fig:Etot}). Both the fermionic and bosonic Green function exhibit a clear light-cone effect after the quench, which is due to the propagation of entangled pairs of quasiparticles.\cite{CalabreseCardy06} However, as compared to the sudden quench the light cone lags behind (see Fig.~\ref{fig:GFcontour1}), which can be associated with a combination of two effects: First, during the quench the quasiparticles propagate at the instantaneous velocity, which is generically smaller than the velocity after the quench. Second, the creation of the quasiparticles is spread over the whole quench duration, while in the sudden case all quasiparticles are created at $t=0$. For short to moderate lengths of the quench we obtain an analytic result for the observed lag, linking it to the integrated change in the coupling during the quench [see Eq.~\eqref{eq:lagapprox}].

The outline of this paper is as follows. In Sec.~\ref{sec:model} we define the Tomonaga--Luttinger model (TLM) and discuss its relation to microscopic fermionic and bosonic systems. In Sec.~\ref{sec:quenches} we present the general approach to the problem of time-dependent interaction quenches in the TLM, and derive some universal properties of the solutions. In Sec.~\ref{sec:exact} we present exact analytic results for several quench protocols including the linear ramp, the smooth cosine quench and periodic quenches with arbitrary number of oscillations. In Sec.~\ref{sec:results} we analyse the behaviour of the total and kinetic energies as well as the fermionic and bosonic Green functions, both during and after the quench. We conclude with a brief discussion of our results in Sec.~\ref{sec:conclusion}. Some technical aspects are presented in the appendices.

\section{Tomonaga--Luttinger model}\label{sec:model}
\subsection{The model}\label{sec:b-model}
In this article we consider the time-dependent Tomonaga--Luttinger model (TLM)\cite{Giamarchi04,Cazalilla-11,Schoenhammer13} defined by
the Hamiltonian
\begin{equation}
\begin{split}
 H(t)  = \sum_{n >0} q_n &\left[\left( v_\text{F} + \frac{g_4(q_n,t)}{2 \pi} \right)
\left( b_n^\dag b_n^{} + b_{-n}^\dag b_{-n}^{} \right) \right. \\
&\qquad\left.  +   \frac{g_2(q_n,t)}{2 \pi}
\left( b_n^\dag b_{-n}^\dag + b_{-n}^{} b_{n}^{} \right)  \right] ,
\label{eq:TLM}
\end{split}
\end{equation}
where $q_n=2 \pi n / L$, $n \in {\mathbb Z}$, and $L$ denote the momenta and system length, respectively, and $v_\text{F}$ is the Fermi velocity. The operators $b_n^\dagger$ and $b_n$ create and annihilate bosonic modes at momentum $q_n$ and  satisfy the standard commutation relations $\comm{b_m}{b_n^\dagger}=\delta_{mn}$. Here and in the following we denote quantities taken at momenta $q_n$ by the subindex $n$. 

Before we proceed with the analysis of the dynamics in the TLM, we briefly recall the properties of the time-independent system given by \eqref{eq:TLM} with coupling functions $g_2(q_n)$ and $g_4(q_n)$ constant in time. Then the Hamiltonian can be diagonalised to $H = \sum_{n \neq 0} \epsilon(q_n) \,\alpha_n^\dagger \alpha_n  + E_{\rm gs}$ by introducing new modes $\alpha_n = c(q_n) b_n + s(q_n) b^\dagger_{-n}$ with
\begin{eqnarray}
s(q)^2 &=& \frac{1}{2} \left[ \frac{1+ \hat{g}_4(q)}{W(q)} -1  \right]  = c(q)^2 -1,\label{eq:defsc}\\
\epsilon(q)&=& v_\text{F}  |q| \, W(q) = v_\text{F}  |q|\sqrt{[1+ \hat{g}_4(q)]^2 - \hat{g}_2(q)^2},\label{eq:epsilon}\quad
\end{eqnarray}
where $\hat{g}_{2/4}(q)= g_{2/4}(q)/(2 \pi v_{\rm F})$ denote dimensionless coupling functions and $E_{\rm gs}=v_\text{F}\sum_{n>0}q_n[W(q_n)-\hat{g}_4(q_n)-1]$ is the ground-state energy. In this article we assume that $g_2(q)$ and $g_4(q)$ depend on the momentum in the dimensionless combination $q/q_\text{c}$ and fall off to zero at $q/q_\text{c}\sim 1$. Thus the scale $q_\text{c}$ provides an ultra-violet cutoff for the theory. Furthermore, we require the limits $\lim_{q\to 0} g_{2/4}(q)$ to be smooth, corresponding to systems with interactions of finite range in real space. Apart from this the momentum dependence is kept arbitrary. In fact, it can be shown\cite{Solyom79,Meden99} that in equilibrium the momentum dependence of $g_2(q)$ and $g_4(q)$ is irrelevant in a renormalisation-group sense, ie, the behaviour at low energies and long wave lengths is governed solely by their values at $q=0$ through the Luttinger-liquid parameter and the renormalised velocity given by
\begin{equation} 
K= \sqrt{\frac{1+\hat g_4(0) -\hat g_2(0)}{1+\hat g_4(0) +\hat g_2(0)}}, \quad 
v=\frac{\dd\epsilon}{\dd q}\bigg|_{q=0}=v_{\rm F} W(0).
\label{eq:LLparameter}
\end{equation}
For non-interacting systems, $g_2(q)=g_4(q)=0$, this simplifies to $K=1$ and $v=v_\text{F}$. In general, if the system possesses Galilean invariance the product $vK$ is independent of the interaction parameters,\cite{Haldane81prl,Schoenhammer13} implying in turn $g_2(0)=g_4(0)$. 

Throughout this article we focus on universal quantities in the sense that they only depend on $g_2(0)$ and $g_4(0)$ and thus the Luttinger parameter $K$ and the renormalised velocity $v$ given in \eqref{eq:LLparameter}. Unless stated otherwise, the numerical results shown in the plots, eg, Fig.~\ref{fig:GF}, were obtained for the specific choice $g_2(q,\tau)=g_4(q,\tau)=g_0\,\exp[-(q/q_\text{c})^2/2]$.

\subsection{Relation to fermionic systems}\label{sec:fermionicmodel}
As is well known, the TLM describes the low-energy physics of one-dimensional fermionic systems in the absence of an energy gap.\cite{Giamarchi04,Schoenhammer13} For example, one can start from the lattice model of spinless fermions, 
\begin{equation}
\begin{split}
  H_\text{F}=& -t\sum_i \bigl(c^{\dag}_{i} c_{i+1} +c^{\dag}_{i+1} c_{i}\bigr)\\
  &+U\sum_i \left(c^{\dag}_{i} c_i-\frac{1}{2}\right)\left(c^{\dag}_{i+1} c_{i+1}-\frac{1}{2}\right), 
  \end{split}
  \label{eq:spinless-ferm-def}
\end{equation}
where $c^{\dag}_{i}$ and $c_i$ are the fermionic creation and annihiliation operators at lattice site $i$. The low-energy description in the gapless regime $U\le 2t$ is obtained by first linearising the dispersion relation around the Fermi points $\pm q_\text{F}$ and taking the continuum limit,
\begin{equation}
\frac{c_i}{\sqrt{a_0}}\to \Psi_\text{F}(x)=e^{\ii q_\text{F}x}\,\Psi_+(x)+e^{-\ii q_\text{F} x}\,\Psi_-(x),
\end{equation}
where $a_0$ denotes the lattice spacing. The right- and left-moving fermionic fields $\Psi_\pm(x)$ are slowly varying on the scale $1/q_\text{F}$. The low-energy properties of \eqref{eq:spinless-ferm-def} are then captured by the Hamiltonian
\begin{equation}
\begin{split}
H_\text{F}=&-\ii v_\text{F}\int\dd x\,\Bigl[\Psi_+\partial_x\Psi_+-\Psi_-\partial_x\Psi_-\Bigr]\\
&+\int\dd x\,\dd x'\,\Bigl[g_4(x-x') \rho_{\pm}(x)\rho_{\pm}(x')\\
&\qquad\qquad\qquad+ g_2(x-x') \rho_{\pm}(x)\rho_{\mp}(x')\Bigr],
\end{split}
\end{equation}
where $\rho_{\pm}(x)=\Psi_{\pm}^\dag(x) \Psi_{\pm}(x)$ are the right- and left-moving densities, and $g_2(x-x')$ and $g_4(x-x')$, the Fourier transforms of $g_2(q)$ and $g_4(q)$, respectively, have thus a straightforward interpretation as density-density couplings.

In the continuum limit one can introduce phase fields corresponding to collective excitations in the electronic liquid via\cite{Schoenhammer13} 
\begin{equation}
\Psi_+(x)=O_+\,e^{\ii\phi^\dagger(x)}\,e^{\ii\phi(x)},\quad
\ii\phi(x)=\sum_{n>0}\frac{e^{\ii q_nx}}{\sqrt{n}}b_n,
\label{eq:RMtomodes}
\end{equation}
where the Klein factor $O_+$ lowers the fermion number by one and commutes with the bosonic modes. The relation between the microscopic parameters $t$ and $U$ of the lattice model and the effective parameters $K$ and $v$ in the TLM is exactly known\cite{Giamarchi04} from the Bethe-ansatz solution of \eqref{eq:spinless-ferm-def}
\begin{equation}
K=\frac{\pi}{2(\pi-\arccos\eta)},\quad v=\frac{\pi t\sqrt{1-\eta^2}}{\arccos\eta},\quad \eta=\frac{U}{2t}.
\end{equation}
We note that the relation between the microscopic and effective parameters is non-linear, implying, for example, that linear time dependences of the former will lead to non-linear quench protocols for the latter. The time evolution of the kinetic energy and fermionic quasiparticle weight (see Sec.s~\ref{sec:kineticenergy} and~\ref{sec:GF} below) as well as local densities after a sudden interaction quench in the lattice model \eqref{eq:spinless-ferm-def} have been analysed\cite{KRSM12,KennesMeden13,HamerlaUhrig13} in the framework of the TLM. 

Since one-dimensional spin models like the XXZ Heisenberg chain can be mapped to fermionic chains of the form \eqref{eq:spinless-ferm-def}, the results presented in our paper can be applied to the analysis of the time evolution during and after finite-time quenches in spin chains. A similar analysis has been performed for the dynamics of several observables in the XXZ chain after sudden quenches\cite{DeChiara-06,Barmettler-09,Barmettler-10,Coira-13,Collura-15} as well as during linear ramps in the anisotropy.\cite{Pollmann-13}

\subsection{Relation to bosonic systems}\label{sec:bosonicmodel}
The TLM also appears in the description\cite{Giamarchi04,Cazalilla-11} of one-dimensional bosonic systems. For example, one may start from the continuum model
\begin{equation}
\begin{split}
H_\text{B}=&\frac{1}{2m_\text{B}}\int\dd x \big|\partial_x\Psi_\text{B}\big|^2-\mu\int\dd x\,\rho_\text{B}(x)\\
&+\frac{1}{2}\int\dd x\,\dd x'\,V(x-x')\rho_\text{B}(x)\rho_\text{B}(x'),
\end{split}
\label{eq:Hboson}
\end{equation}
were $\Psi_\text{B}(x)$ is a complex scalar field describing bosons of mass $m_\text{B}$, $\rho_\text{B}(x)=\Psi^\dagger_\text{B}(x)\Psi_\text{B}(x)$ denotes the boson density, $V(x)$ is a density-density interaction, and $\mu$ the chemical potential. In the special case $V(x)=V_0\delta(x)$ the model becomes the integrable Lieb--Liniger model,\cite{LiebLiniger63} whose Bethe-ansatz solution can be used to study its time evolution after a sudden quench.\cite{Gritsev-10,IyerAndrei12,Iyer-13,DeNardis-14,DeNardisCaux14,Zill-15,DeNardis-15,vandenBerg-16} In the presence of a lattice a natural starting point would be the Bose--Hubbard model
\begin{equation}
\begin{split}
H_\text{BHM}=&-t\sum_i \bigl(a^{\dag}_{i} a_{i+1} +a^{\dag}_{i+1} a_{i}\bigr)\\
&+U\sum_i n_i(n_i-1)-\mu\sum_i n_i, 
\end{split}
\label{eq:BHM}
\end{equation}
where $a^{\dag}_{i}$ and $a_i$ are the bosonic creation and annihiliation operators at lattice site $i$ and $n_i=a_i^\dagger a_i$ denotes the respective density. We note in passing that models like \eqref{eq:Hboson} and \eqref{eq:BHM} can be realised in systems of trapped, ultra-cold atoms.\cite{Buechler-03} In the superfluid phase the low-energy properties of bosonic systems like \eqref{eq:Hboson} and \eqref{eq:BHM} can be described by the TLM \eqref{eq:TLM}. For that one writes the bosonic fields in terms of the density $\rho_\text{B}$ and phase field $\theta$ as\cite{Cazalilla-11} 
\begin{equation}
\Psi_\text{B}(x)=\sqrt{\rho_\text{B}(x)}\,e^{\ii\theta(x)},
\end{equation}
where the phase field can in turn be represented by the creation and annihilation operators of the bosonic modes as
\begin{equation}\label{eq:phi-to-b}
  \theta(x)=\frac{\ii}{2}\sum_{n\neq 0} \frac{\exp(-\ii q_n x)}{\sqrt{n}}\bigl(b_n^{\dag}-b_{-n}\bigr).
\end{equation}
The effective parameters $K$ and $v$ in the TLM can be obtained for example from the Bethe-ansatz solution in the Lieb--Liniger case or via the density-matrix renormalisation group method from the Bose--Hubbard model.\cite{Kollath-04} Again the relation between the microscopic and effective parameters is non-linear. The time evolution of the bosonic systems \eqref{eq:Hboson} and \eqref{eq:BHM} during finite-time interaction quenches was investigated in Refs.~\onlinecite{Bernier-12,HaqueZimmer13,Bernier-14}.

\section{Finite-time quench protocols}\label{sec:quenches}
After discussing the basic properties of the TLM and its relation to other one-dimensional systems, we now turn to the time evolution due to a change in the coupling functions in \eqref{eq:TLM}. Starting with the seminal work by Cazalilla\cite{Cazalilla06} the dynamics of the TLM after a sudden quench, ie, a sudden change in the coupling functions, has been exhaustively studied in the past.\cite{Perfetto06,DeChiara-06,Barmettler-09,Uhrig09,IucciCazalilla09,Barmettler-10,MitraGiamarchi11,Mitra12,KRSM12,MitraGiamarchi12,RSM12,DallaTorre-13,Mitra13,NessiIucci13,KennesMeden13,NgoDinh-13,HamerlaUhrig13,Coira-13,Tavora-14,Kennes-14,Protopopov-15,Collura-15}

\begin{figure}[t]
    \centering
    \includegraphics[width=0.49\textwidth]{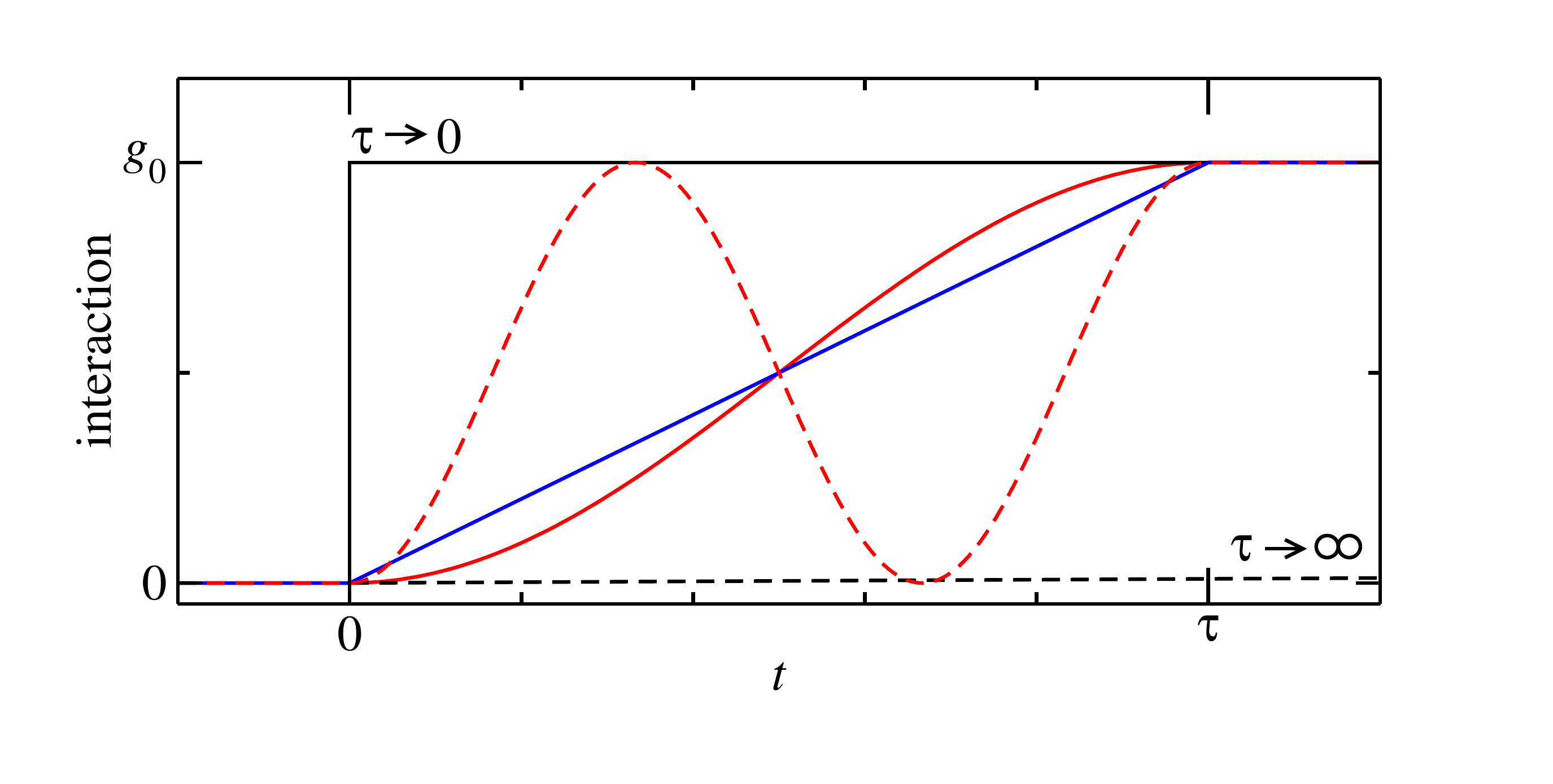}
    \caption{(Colour online) Sketch of different time-dependent protocols for a general, finite-time quench. Here $\tau$ denotes the quench time over which the parameter $g$ varies. The solid and dashed black lines represent the special cases of sudden $(\tau\to 0)$ and adiabatic $(\tau\to\infty)$ quenches respectively. The solid blue line corresponds to a linear ramp \eqref{eq:linearquench}, which constitutes the simplest finite-time quench. Further examples considered here are the cosine quench \eqref{eq:cosinequench} and the periodic quench \eqref{eq:cosine3quench} illustrated by the solid and dashed red lines respectively.}
    \label{fig:nonsudden}
\end{figure}
In contrast, here we will consider the dynamics during and after continuous changes in the coupling functions; see Fig.~\ref{fig:nonsudden} for an illustration. Most previous works investigating such a setup have focused on linear quenches in which the interactions are ramped up at a constant speed.\cite{Dora-11,Pollmann-13,Bernier-14,Sachdeva-14} We consider quench protocols starting from the non-interacting model and ranging over a finite quench time $\tau$, ie, the coupling functions satisfy
\begin{equation}
g_{2/4}(q,t<0)=0, \quad g_{2/4}(q,t>\tau)=g_{2/4}(q),
\end{equation}
with $g_{2/4}(q)$ being independent of time. At zero temperature, to which we restrict ourselves, the initial state $\ket{\Psi_0}$ at $t=0$ is thus the vacuum state of the bosons, ie, $b_n\ket{\Psi_0}=0$ for all $n$.

\subsection{Evolution during the quench, $\boldsymbol{t<\tau}$}\label{sec:during}
The time evolution of the bosonic operators is governed by the Heisenberg equations of motion\cite{Dora-11}
\begin{equation}
\ii\frac{\dd}{\dd t}O(t)=\bigl[O(t),H(t)\bigr],\quad O(t)=b_n(t), b_n^\dagger(t),
\end{equation}
where we denote operators in the Heisenberg picture by stating the time argument explicitly. For the TLM we obtain
\begin{eqnarray}
\ii\frac{\dd}{\dd t}b_n(t)&=&\omega_n(t)b_n(t)+\lambda_n(t)b_{-n}^\dagger(t),\label{eq:Heisenbergb1}\\
\ii\frac{\dd}{\dd t}b_n^\dagger(t)&=&-\omega_n(t)b_n^\dagger(t)-\lambda_n(t)b_{-n}(t),\label{eq:Heisenbergb2}
\end{eqnarray}
with the abbreviations 
\begin{eqnarray}
\omega(q,t)&=&v_\text{F}|q|\bigl[1+\hat{g}_4(q,t)\bigr]=v_\text{F}|q|\left(1+\frac{g_4(q,t)}{2\pi v_\text{F}}\right)\!,\quad\\
\lambda(q,t)&=&v_\text{F}|q|\hat{g}_2(q,t)=\frac{|q|}{2\pi}g_2(q,t),
\end{eqnarray}
and $\omega_n(t)=\omega(q_n,t)$ as well as $\lambda_n(t)=\lambda(q_n,t)$. Here and in the following we denote quantities taken at momenta $q_n$ by the subindex $n$. Now making the ansatz
\begin{eqnarray}
b_n(t)&=&u_n(t)b_n+v_n(t)^*\,b_{-n}^\dagger,\\
b_n^\dagger(t)&=&u_n(t)^*\,b_n^\dagger+v_n(t)b_{-n},
\end{eqnarray}
where the operators on the right-hand side are the time-independent Schr\"odinger operators at $t=0$, the equations \eqref{eq:Heisenbergb1} and \eqref{eq:Heisenbergb2} turn into differential equations for the coefficients\cite{Dora-11} $u_n(t)$ and $v_n(t)$,
\begin{equation}
\ii\frac{\dd}{\dd t}\left(\begin{array}{c}u_n(t)\\ v_n(t)\end{array}\right)=
\left(\begin{array}{cc}\omega_n(t)& \lambda_n(t)\\ -\lambda_n(t) & -\omega_n(t)\end{array}\right)
\left(\begin{array}{c}u_n(t)\\ v_n(t)\end{array}\right).
\label{eq:DGLuv}
\end{equation}
The initial conditions read
\begin{equation}
u_n(0)=1,\quad v_n(0)=0,
\label{eq:uvinit}
\end{equation}
and we recall $u_n(t)=u(q_n,t)$ and $v_n(t)=v(q_n,t)$. The coefficients satisfy $u(q,t)=u(-q,t)$, $v(q,t)=v(-q,t)$ and $|u(q,t)|^2-|v(q,t)|^2=1$ as well as $\lim_{\tau\to 0}u_n(t=\tau)=1$ and $\lim_{\tau\to 0}v_n(t=\tau)=0$.

Furthermore, since the universal properties of the system are governed by the behaviour at small momenta, we determine the expansion of the solutions $u(q,t)$ and $v(q,t)$ at $q\ll q_\text{c}$. More specifically we make the ansatz
\begin{equation}
u(q,t)=\sum_{m=0}^\infty u^{(m)}(t)\,\left(\frac{q}{q_\text{c}}\right)^m,
\end{equation}
and similarly for $v(q,t)$, $\omega(q,t)$ and $\lambda(q,t)$. [We recall that the coupling functions $g_{2/4}(q,t)$ are assumed to be analytic at $q=0$, thereby excluding long-range interactions in real space.] Then we obtain for the initial conditions \eqref{eq:uvinit} and up to linear order in $q$ the results 
\begin{eqnarray}
u(q,t)&=&1-\ii v_\text{F}q\int_0^t\dd t'\,\bigl[1+\hat{g}_4(0,t')\bigr],\label{eq:smallku}\\
v(q,t)&=&\ii v_\text{F}q\int_0^t\dd t'\,\hat{g}_2(0,t').\label{eq:smallkv}
\end{eqnarray}
These expansions are valid for sufficiently small momenta $q\ll q_\text{c},1/(v_\text{F}\tau)/[1+\hat{g}_4(0)]\sim 1/(v_\text{F}\tau)$, where the second condition originates from the requirement that the next-to-leading order term in \eqref{eq:smallku} stays smaller than the leading one (see App.~\ref{app:smallk} for a more detailed discussion). Apart from the restriction on $q$ the expansions \eqref{eq:smallku} and \eqref{eq:smallkv} are valid for quench protocols with arbitrary time dependencies and final interaction strengths $g_{2/4}(0,\tau)$. In particular, they are applicable in the non-perturbative regime $1\lesssim\hat{g}_{2/4}(0,\tau)$. It is straightforward to explicitly verify \eqref{eq:smallku} and \eqref{eq:smallkv} for the exactly solvable protocols discussed in Sec.~\ref{sec:exact} [but also the perturbative solution of \eqref{eq:DGLuv} presented for completeness in App.~\ref{app:PT}].

\subsection{Evolution after the quench, $\boldsymbol{t>\tau}$}\label{sec:after}
For times $t>\tau$ the coupling functions are constant and thus the differential equations \eqref{eq:DGLuv} can be solved explicitly by
\begin{equation}
v_n(t)=A_n\cos(\epsilon_nt)+B_n\sin(\epsilon_nt),
\label{eq:vafterquench}
\end{equation}
where [we assume $\omega_n(\tau)>\lambda_n(\tau)$]
\begin{equation}
\epsilon_n=\epsilon(q_n)=\sqrt{\omega_n(\tau)^2-\lambda_n(\tau)^2}>0
\end{equation}
is the single-mode energy \eqref{eq:epsilon} after the quench, and 
\begin{equation}
u_n(t)=-\frac{\ii}{\lambda_n(\tau)}\frac{\dd}{\dd t}v_n(t)-\frac{\omega_n(\tau)}{\lambda_n(\tau)}v_n(t).
\end{equation}
The constants $A_n=A(q_n)$ and $B_n=B(q_n)$ are obtained from the initial conditions for the post-quench dynamics at $t=\tau$,
\begin{eqnarray}
A_n&=&-\ii\frac{\lambda_n(\tau)}{\epsilon_n}\sin(\epsilon_n\tau)u_n(\tau)\\*
& &-\frac{\ii}{\epsilon_n}\bigl[\ii\epsilon_n\cos(\epsilon_n\tau)+\omega_n(\tau)\sin(\epsilon_n\tau)\bigr]v_n(\tau),\nonumber\\
B_n&=&\ii\frac{\lambda_n(\tau)}{\epsilon_n}\cos(\epsilon_n\tau)u_n(\tau)\\*
& &-\frac{\ii}{\epsilon_n}\bigl[\ii\epsilon_n\sin(\epsilon_n\tau)-\omega_n(\tau)\cos(\epsilon_n\tau)\bigr]v_n(\tau).\nonumber
\end{eqnarray}
The expansion of the post-quench coefficients for small momenta is obtained using \eqref{eq:smallku} and \eqref{eq:smallkv}; to leading order it reads
\begin{eqnarray}
A(q)&=& -\ii v_\text{F}q\tau\left[\hat{g}_2(0,\tau)-\frac{1}{\tau}\int_0^\tau\dd t\,\hat{g}_2(0,t)\right],\label{eq:smallkA}\\
B(q)&=&\ii\frac{1-K^2}{2K},\label{eq:smallkB}
\end{eqnarray}
with the Luttinger parameter $K$ defined in \eqref{eq:LLparameter}. The expansion is valid for arbitrary quench protocols provided $q\ll q_\text{c},1/(v_\text{F}\tau)$. For linear or periodic quenches (see next section for the precise definition) the expansion \eqref{eq:smallkA} simplifies to $A(q)=-\ii v_\text{F}\hat{g}_2(q,\tau) q\tau/2$.

\subsection{Generalised Gibbs ensemble}\label{sec:GGE}
\begin{figure}[t]
	\centering
	\includegraphics[width=0.49\textwidth]{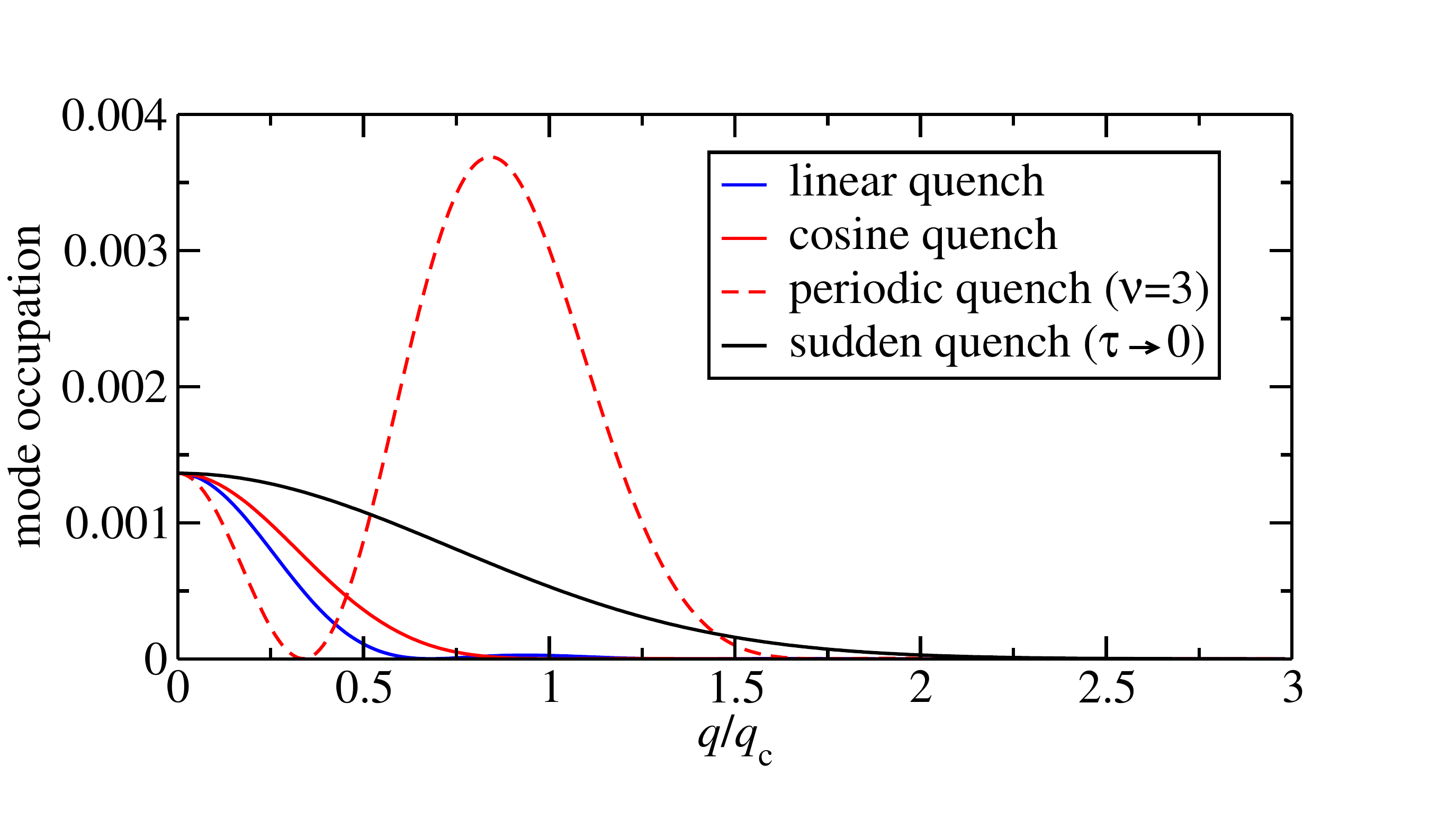}
	\caption{(Colour online) Mode occupations $\bra{\Psi(\tau)}\alpha_n^\dagger\alpha_n\ket{\Psi(\tau)}$ after the quench with the quench time given by $\ln(v_\text{F}q_\text{c}\tau)=1.5$ (except for the sudden quench), Gaussian momentum dependence of the coupling functions and final interaction strength $g_0=v_\text{F}/2$. For the periodic quench we observe a preferred occupation of high-energy modes  which strongly affects the behaviour of the total energy shown in Fig.~\ref{fig:Etot}. In the adiabatic limit $\tau\to\infty$ one finds $\bra{\Psi(\tau)}\alpha_n^\dagger\alpha_n\ket{\Psi(\tau)}\to 0$ for all $n\neq 0$ and all quench protocols.}
	\label{fig:modes}
\end{figure}
Since the time evolution for $t>\tau$ is governed by the time-independent Hamiltonian \eqref{eq:TLM} conserving the mode occupations after the quench, the system relaxes for $t\to\infty$ to a generalised Gibbs ensemble.\cite{Rigol-07} One obtains\cite{Dora-12} (in complete analogy to the situation after a sudden quench\cite{Cazalilla06,IucciCazalilla09})
\begin{equation}
\rho_\text{GGE}=\frac{e^{-\sum_{n\neq 0}\eta_n\alpha_n^\dagger \alpha_n}}{\text{tr}\left(e^{-\sum_{n\neq 0}\eta_n\alpha_n^\dagger \alpha_n}\right)},
\label{eq:GGE}
\end{equation}
where $\alpha_n^\dagger\alpha_n$ are the mode-occupation operators after the quench and the Lagrange multipliers are obtained from the initial conditions for the post-quench dynamics. Explicitly we find with the coefficients defined in \eqref{eq:defsc},
\begin{equation}
\begin{split}
&\bra{\Psi(\tau)}\alpha_n^\dagger\alpha_n\ket{\Psi(\tau)}=\big|c(q_n)v_n(\tau)+s(q_n)u_n(\tau)\big|^2\\
&\qquad\stackrel{!}{=}\text{tr}\bigl(\rho_\text{GGE}\alpha_n^\dagger\alpha_n\bigr)=\frac{e^{-\eta_n}}{1-e^{-\eta_n}}.
\end{split}
\label{eq:modesGGE}
\end{equation}
We stress that the Lagrange multipliers implicitly depend on the quench time $\tau$ and the precise form of the time dependence for $0<t<\tau$ via the coefficients $u_n(\tau)$ and $v_n(\tau)$. In the sudden limit we recover the well-known\cite{Cazalilla06,IucciCazalilla09} result $\bra{\Psi(\tau)}\alpha_n^\dagger\alpha_n\ket{\Psi(\tau)}=s(q_n)^2$. 

The mode occupations \eqref{eq:modesGGE} for specific quench protocols are illustrated in Fig.~\ref{fig:modes}. In particular, we observe that the periodic quench leads to a preferred occupation of high-energy modes around the band edge $q\sim q_\text{c}$ for quench times $\tau\sim 5/(v_\text{F}q_\text{c})$, ie, when the energy scale $1/\tau$ related to the quench protocol is comparable to the band width. In contrast, for the sudden or monotonic quench protocols such a preferred occupation of high-energy modes is not observed. We further note that the generalised Gibbs ensemble \eqref{eq:GGE} is diagonal in the modes and thus does not capture correlations between the $q$ and $-q$ modes.\cite{IucciCazalilla09} This limitation can be overcome by additionally including $\alpha_n^\dagger\alpha_n\alpha_{-n}^\dagger\alpha_{-n}$ into the set of conserved quantities used to define the generalised Gibbs ensemble.

\section{Analytically solvable quench protocols}\label{sec:exact}
In the case of Galilean invariance, $g_2(q,t)=g_4(q,t)$, exact solutions are possible for specific time dependences. (We briefly comment on the general case in Sec.~\ref{sec:non-Galilean-def}.) In order to derive them we introduce an auxiliary function $a_n(t)$ via
\begin{eqnarray}
u_n(t)&=&\frac{1}{2}a_n(t)+\frac{\ii}{2v_\text{F}|q_n|}\frac{\dd}{\dd t}a_n(t),\label{eq:relationua}\\
v_n(t)&=&\frac{1}{2}a_n(t)-\frac{\ii}{2v_\text{F}|q_n|}\frac{\dd}{\dd t}a_n(t).\label{eq:relationva}
\end{eqnarray}
In terms of this auxiliary function the differential equation \eqref{eq:DGLuv} becomes
\begin{equation}
\frac{\dd^2}{\dd t^2}a_n(t)+v_\text{F}^2q_n^2\bigl[1+2\hat{g}_2(q_n,t)\bigr]a_n(t)=0,
\label{eq:DGLa}
\end{equation}
with the initial conditions
\begin{equation}
a_n(0)=1,\quad \frac{\dd}{\dd t}a_n(t)\Big|_{t=0}=-\ii v_\text{F}|q_n|.
\label{eq:DGLainit}
\end{equation}
For specific choices of the time dependence of $g_2(q_n,t)$ the differential equation \eqref{eq:DGLa} admits a closed solution which we discuss in the following subsections.

\subsection{Linear quench}\label{sec:linear}
The simplest finite-time quench protocol is a linear ramp
\begin{equation}
g_2(q,t)=g_4(q,t)=g_2(q)\,\frac{t}{\tau},
\label{eq:linearquench}
\end{equation}
which is sketched by the solid blue line in Fig.~\ref{fig:nonsudden}. In our setup, contrary to most of previous works \cite{Dora-11,PerfettoStefanucci11,DziarmagaTylutki11,Dora-12,Pollmann-13,Sachdeva-14}, we do not neglect the $g_4$-term. We note that the coupling functions in the linear quench are not differentiable at $t=0$ and $t=\tau$. Inserting \eqref{eq:linearquench} into \eqref{eq:DGLa} leads to the Airy differential equation\cite{AbramowitzStegun65} whose solution for the initial conditions \eqref{eq:DGLainit} can be rewritten in terms of Bessel functions as\cite{Bernier-14,footnote1}
\begin{equation}
\begin{split}
a_n(t)=&\frac{\pi h_n\sqrt{\tilde{t}_n}}{\sqrt{3}}\\
&\times\Bigl\{\bigl[J_{2/3}(h_n)-\ii J_{-1/3}(h_n)\bigr]J_{1/3}(h_n\tilde{t}_n^{3/2})\\
&\quad+\bigl[J_{-2/3}(h_n)+\ii J_{1/3}(h_n)\bigr]J_{-1/3}(h_n\tilde{t}_n^{3/2})\Bigr\},
\end{split}
\label{eq:solutionlinearquench}
\end{equation}
where we have introduced $h_n=v_\text{F}|q_n|\tau/[3\hat{g}_2(q_n)]$ and $\tilde{t}_n=1+2\hat{g}_2(q_n)t/\tau$. The solution \eqref{eq:solutionlinearquench} is valid during the quench $0\le t\le \tau$. The time evolution after the quench is given by the results stated in Sec.~\ref{sec:after} with the constants $A_n$ and $B_n$ encoding the evolution for $t<\tau$ via the values of $u_n(\tau)$ and $v_n(\tau)$ obtained from \eqref{eq:solutionlinearquench}.

\subsection{Cosine quench}\label{sec:cosine}
Next we consider a quench protocol where also the first derivative at $t=0$ and $t=\tau$ is continuous. This is for example the case for the cosine quench
\begin{equation}
g_2(q,t)=g_4(q,t)=\frac{g_2(q)}{2}\left[1-\cos\frac{\pi t}{\tau}\right],
\label{eq:cosinequench}
\end{equation}
sketched as solid red line in Fig.~\ref{fig:nonsudden}. The exact solution of the differential equation \eqref{eq:DGLa} with the initial condition \eqref{eq:DGLainit} is given by
\begin{equation}
\begin{split}
a_n(t)=&c_1\,\text{Ce}\biggl(2\tilde{h}_n\bigl[1+\hat{g}_2(q_n)\bigr],\tilde{h}_n\hat{g}_2(q_n),\frac{\pi t}{2\tau}\biggr)\\
&+c_2\,\text{Se}\biggl(2\tilde{h}_n\bigl[1+\hat{g}_2(q_n)\bigr],\tilde{h}_n\hat{g}_2(q_n),\frac{\pi t}{2\tau}\biggr),
\end{split}
\label{eq:solutioncosinequench}
\end{equation}
where $\text{Ce}(a,q,z)$ and $\text{Se}(a,q,z)$ denote the even and odd Mathieu functions\cite{AbramowitzStegun65,footnote2} satisfying the differential equation $y''+[a-2q\cos(2z)]y=0$, $\tilde{h}_n=2(v_\text{F}q_n\tau)^2/\pi^2$, and the integration constants are
\begin{equation}
\begin{split}
\frac{1}{c_1}&=\text{Ce}\bigl(2\tilde{h}_n\bigl[1+\hat{g}_2(q_n)\bigr],\tilde{h}_n\hat{g}_2(q_n),0\bigr),\\
\frac{1}{c_2}&=\frac{\ii\pi}{2v_\text{F}|q_n|\tau}\frac{\partial}{\partial z}\text{Se}\bigl(2\tilde{h}_n\bigl[1+\hat{g}_2(q_n)\bigr],\tilde{h}_n\hat{g}_2(q_n),z\bigr)\Big|_{z=0}.
\end{split}
\end{equation}

\subsection{Periodic quench}\label{sec:periodic}
We note that the solution \eqref{eq:solutioncosinequench} also allows one to consider periodic driving\cite{KaganManakova09,Graf-10,Pielawa11,BukovHeyl12} of the system. In fact, for periodic quenches with any period, ie, quenches of the form
\begin{equation}
g_2(q,t)=g_4(q,t)=\frac{g_2(q)}{2}\left[1-\cos\frac{\nu\pi t}{\tau}\right],\quad\nu\in\mathbb{N},
\label{eq:cosine3quench}
\end{equation}
the solution for $a_n(t)$ during the quench is directly obtained from \eqref{eq:solutioncosinequench} via the replacement $\tau\to\tau/\nu$. For odd $\nu$ these quenches correspond to the switching on of interactions after periodic driving (see the dashed red line in Fig.~\ref{fig:nonsudden} for a sketch), while for even $\nu$ the system is driven for $\nu/2$ periods and then returns to the non-interacting Hamiltonian. In the context of periodically driven Luttinger liquids the solution \eqref{eq:solutioncosinequench} has already been employed\cite{Pielawa11,BukovHeyl12} to investigate features like parametric resonances and metastable states.\cite{KaganManakova09,Graf-10,Pielawa11,BukovHeyl12}

\subsection{Exponential and quadratic quenches}\label{sec:exponential}
It is also possible to treat exponential quenches of the form
\begin{equation}
g_2(k,t)=g_4(k,t)=\frac{g_2(k)}{\xi}\left(e^{t\ln(1+\xi)/\tau}-1\right)
\label{eq:exponentialquench}
\end{equation}
with $\xi>0$, which lead to exact solutions in terms of Bessel functions\cite{AbramowitzStegun65} 
\begin{equation}
a_n(t)=c_1\,J_{-\nu_n}\bigl(\hat{t}_n\bigr)+c_2\,J_{\nu_n}\bigl(\hat{t}_n\bigr),
\end{equation}
where
\begin{eqnarray}
\nu_n&=&\frac{2v_\text{F}|q_n|\tau\sqrt{2\hat{g}_2(q_n)-\xi}}{\sqrt{\xi}\,\ln(1+\xi)},\\
\hat{t}_n&=&\frac{\sqrt{8}v_\text{F}|q_n|\tau\sqrt{\hat{g}_2(q_n)\,(1+\xi)^{t/\tau}}}{\sqrt{\xi}\,\ln(1+\xi)}.
\end{eqnarray}
The integration constants $c_1$ and $c_2$ have to be determined from the initial condition \eqref{eq:DGLainit}. Similarly, quadratic quenches
\begin{equation}
g_2(k,t)=g_4(k,t)=g_2(k)\,\frac{t^2}{\tau^2}
\label{eq:quadraticquench}
\end{equation}
result in solutions in terms of parabolic cylinder functions\cite{AbramowitzStegun65}
\begin{equation}
\begin{split}
a_n(t)=&c_1\,D_{-\frac{1}{2}+\mu_n}\left(-e^{-\ii\frac{\pi}{4}}2^{3/4}\hat{g}_2(q_n)^{1/4}\sqrt{\frac{v_\text{F}q_n}{\tau}}t\right)\\
&+c_2\,D_{-\frac{1}{2}-\mu_n}\left(e^{\ii\frac{\pi}{4}}2^{3/4}\hat{g}_2(q_n)^{1/4}\sqrt{\frac{v_\text{F}q_n}{\tau}}t\right)
\end{split}
\end{equation}
with $\mu_n=\ii v_\text{F}q_n\tau/\sqrt{8\hat{g}_2(q_n)}$.

\subsection{Beyond Galilean invariance}\label{sec:non-Galilean-def}
If we drop the requirement of Galilean invariance, ie, if we allow coupling functions with $g_2(q,t)\neq g_4(q,t)$, a differential equation similar to \eqref{eq:DGLa} can still be derived. Defining $a_n(t)=u_n(t)+v_n(t)$ [we stress that \eqref{eq:relationua} and \eqref{eq:relationva} are no longer valid] and taking the second derivative we obtain
\begin{equation}
\begin{split}
&\ddot{a}_n(t)-\frac{\dot{a}_n(t)}{1+\hat{g}_4(q_n,t)-\hat{g}_2(q_n,t)}\frac{\dd}{\dd t}\bigl[\hat{g}_4(q_n,t)-\hat{g}_2(q_n,t)\bigr]\\
&\quad+v_\text{F}^2q_n^2\Bigl[\bigl(1+\hat{g}_4(q_n,t)\bigr)^2-\hat{g}_2(q_n,t)^2\Bigr]a_n(t)=0,
\end{split}
\label{eq:DGLa-beyon}
\end{equation}
with the initial conditions for non-interacting initial states still given by $a_n(0)=1$ and $\dot{a}_n(0)=-\ii v_\text{F}|q_n|$. However, we are not aware of quench protocols that allow for an exact, analytical solution of \eqref{eq:DGLa-beyon}. The differential equation \eqref{eq:DGLa-beyon} and some properties of its solution will be further investigated in a separate work.\cite{Chudzinski16}

\section{Results for specific observables}\label{sec:results}
In this section we consider the total and kinetic energies and the fermionic and bosonic Green functions. We focus on universal properties that depend only on the values of $g_{2/4}(q)$ at $q=0$, which translates into a dependence on the Luttinger parameter $K$ and renormalised velocity $v$ of the post-quench system. Unless stated otherwise, $g_2(q)$ and $g_4(q)$ are considered to be independent functions. The numerical results shown in the plots of this section were obtained for the specific choice $g_2(q,\tau)=g_4(q,\tau)=g_0\,\exp[-(q/q_\text{c})^2/2]$.

\subsection{Total energy}\label{sec:totalenergy}
\begin{figure}[t]
    \centering
    \includegraphics[width=0.49\textwidth]{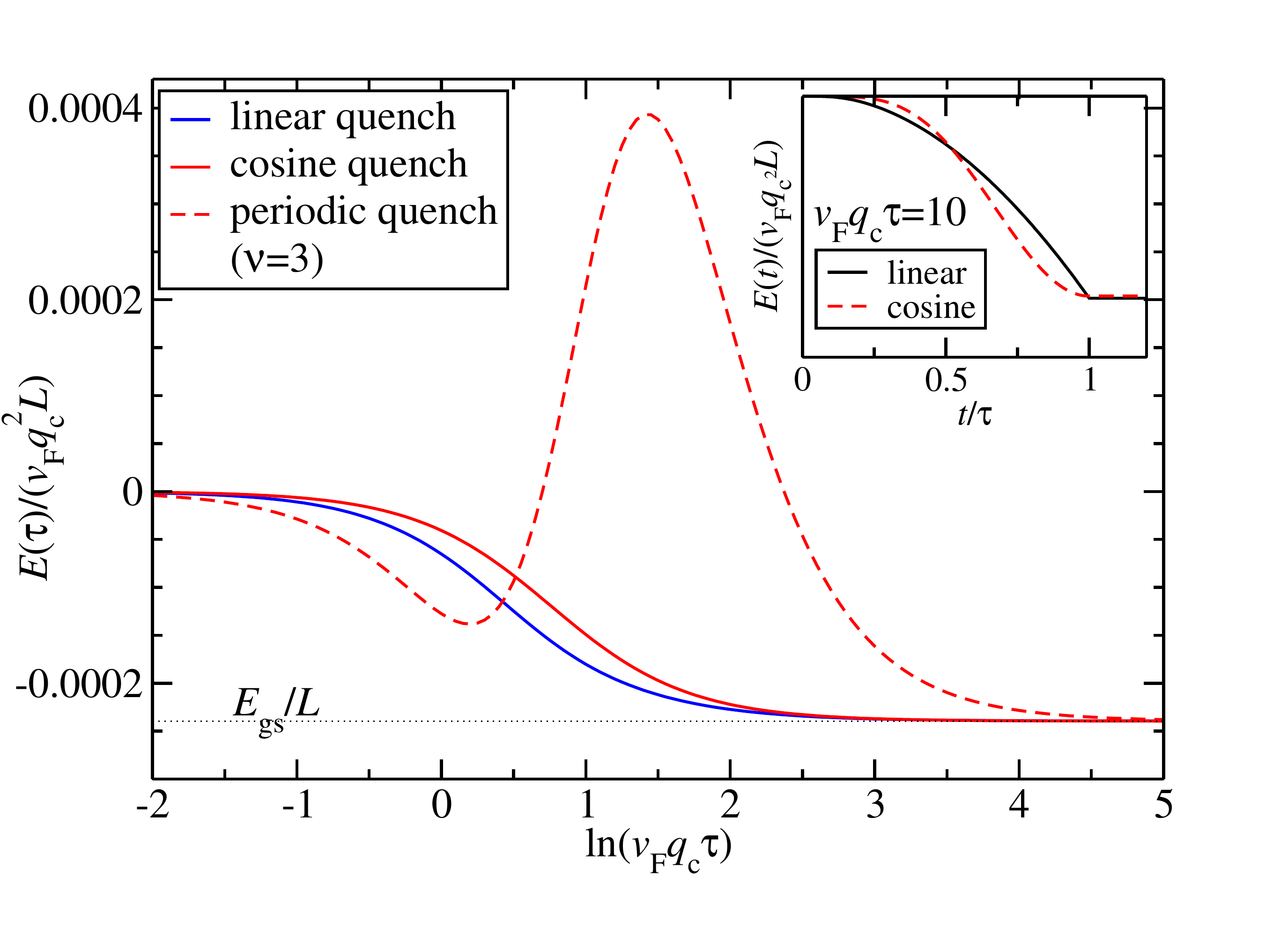}
    \caption{(Colour online) Energy density $E(\tau)/L$ after linear, cosine and periodic quenches with final interaction strength $g_0=v_\text{F}/2$. The time dependence during the quench clearly affects the total energy, in particular for the case with periodic driving for one period. In the adiabatic limit $\tau\to\infty$ the total energy density reaches the ground-state energy density $E_\text{gs}/L$ indicated by the dotted line. Inset: Energy density during linear and cosine quenches of length $v_\text{F}q_\text{c}\tau=10$, the former showing a kink at $t=\tau$.}
    \label{fig:Etot}
\end{figure}
The simplest observable is the total energy $E(t)=\langle H\rangle(t)$ during the quench, which has first been analysed for linear quenches in Ref.~\onlinecite{Dora-11}. Following this work a comparison to numerical data for linear quenches in the gapless phase of the XXZ chain showed very good agreement,\cite{Pollmann-13} thus indicating that the TLM can describe finite-time quenches in lattice models.

The total energy in the TLM reads
\begin{equation}
E(t)=\sum_{n\neq 0}\text{Im}\left[v_n(t)^*\,\frac{\dd}{\dd t}v_n(t)\right],
\label{eq:energy}
\end{equation}
after the quench, $t>\tau$, the total energy is given by $E(\tau)=\text{tr}(\rho_\text{GGE}H)$ with the generalised Gibbs ensemble $\rho_\text{GGE}$ defined in Sec.~\ref{sec:GGE}. The energy \eqref{eq:energy} depends on the details of the quench protocol and the quench time (as well as the precise form of the coupling functions), for example
\begin{equation}
\begin{split}
\frac{\dd}{\dd t}E(t)=&\frac{1}{2\pi}\sum_{n\neq 0}|q_n|\,\Bigl[\dot{g}_2(q_n,t)\,\text{Re}\bigl(v_n(t)^*\,u_n(t)\bigr)\\
&\qquad\qquad\qquad+\dot{g}_4(q_n,t)\,\big|v_n(t)\big|^2\Bigr],
\end{split}
\end{equation}
where the dot denotes the derivative with respect to time. In particular, kinks in $g_{2/4}(q,t)$ result in kinks in the total energy as exemplified in the inset of Fig.~\ref{fig:Etot}. As further shown in Fig.~\ref{fig:Etot} we observe that in both the sudden and the adiabatic limit the result does not depend on the quench protocol, as is of course well expected. In contrast, for quench times of the order of the inverse band width, $\tau\sim 1/(v_\text{F}q_\text{c})$, the results for the linear and cosine quenches clearly differ. The most drastic effect of the finite quench time shows up for the periodic quench, where at quench times $\tau\sim 5/(v_\text{F}q_\text{c})$ an increase of the energy can be observed. The physical origin of this behaviour lies in the preferred occupation of high-energy modes by these quench protocols, as can be seen from the mode occupation after the quench shown in Fig.~\ref{fig:modes}. For the linear and cosine quenches we observe that the adiabatic limit is reached as $E(\tau)-E_\text{gs}\propto (v_\text{F}q_\text{c}\tau)^{-2}\,\ln(v_\text{F}q_\text{c}\tau)$ in accord with the so-called analytic regime discussed in Ref.~\onlinecite{PolkovnikovGritsev08}. The behaviour $E(\tau)-E_\text{gs}\sim\tau^{-2}$ was also observed\cite{HaqueZimmer13} after power-law quenches in various many-body systems in harmonic traps, including the one-dimensional Bose--Hubbard model. Our results for the cosine quench suggest that this general behaviour originates from the existence of a finite quench time $\tau$ rather than from the existence of an endpoint kink at $t=\tau$ in the interaction function.

\subsection{Kinetic energy}\label{sec:kineticenergy}
As first observable which shows non-trivial behaviour after the quench we consider the kinetic energy, which is defined as the expectation value of the non-interacting Hamiltonian $H_\text{kin}=H(t<0)=v_\text{F}\sum_{n\neq 0}|q_n|b_n^\dagger b_n$. Straightforward calculation gives
\begin{equation}
E_\text{kin}(t)=2v_\text{F}\sum_{n>0}q_n\,\big|v_n(t)\big|^2.
\end{equation}
For times $t>\tau$ and in the thermodynamic limit we obtain
\begin{equation}
\begin{split}
E_\text{kin}(t)=&\frac{v_\text{F}L}{\pi}\int_0^\infty \dd q\,q\biggl[\frac{|A(q)|^2+|B(q)|^2}{2}\\*
&\quad+\frac{|A(q)|^2-|B(q)|^2}{2}\cos[2\epsilon(q)t]\\*
&\quad+\text{Re}\bigl[A(q)^*B(q)\bigr]\,\sin[2\epsilon(q)t]\biggr].
\end{split}
\end{equation}
The behaviour at late times can be obtained using asymptotic analysis;\cite{BenderOrszag99} assuming $\dd\epsilon(q)/\dd q\neq 0$ we find
\begin{equation}
E_\text{kin}(t)=E_\text{kin}^\infty+\frac{L\gamma_\text{kin}}{t^2}+\mathcal{O}(t^{-3}).
\label{eq:Ekinlatetimes}
\end{equation}
Here
\begin{equation}
\begin{split}
E_\text{kin}^\infty&=\frac{v_\text{F}L}{2\pi}\int_0^\infty \dd q\,q\,\bigl(|A(q)|^2+|B(q)|^2\bigr)\\
&=\text{tr}\bigl(\rho_\text{GGE}H_\text{kin}\bigr)
\end{split}
\end{equation}
denotes the asymptotic limit identical to the expectation value of $H_\text{kin}$ in the generalised Gibbs ensemble \eqref{eq:GGE}. As such it inherits all properties of the mode occupation after the quench \eqref{eq:modesGGE}, for example the non-monotonic dependence on the quench time after a periodic quench. Furthermore, the decay parameter $\gamma_\text{kin}$ is given by
\begin{equation}
\gamma_\text{kin}=\frac{v_\text{F}}{32\pi v^2}\left(K-\frac{1}{K}\right)^2,
\end{equation}
where $v$ and $K$ are defined in \eqref{eq:LLparameter} with the coupling functions taken at the quench time $t=\tau$. We stress that the decay parameter is universal in the sense that it depends on the values of the coupling functions $g_2(q,\tau)$ and $g_4(q,\tau)$ at $q=0$ only. Moreover, $\gamma_\text{kin}$ is identical to the result for the sudden quench,\cite{KRSM12} ie, it does not depend on the quench time or the precise form of the quench protocol but only on the final values of the interaction. Finally, we note that the derivation of \eqref{eq:Ekinlatetimes} relies on the assumption that $g_2(q,\tau)$ and $g_4(q,\tau)$ are smooth functions of $q$. For example, if we consider a sharp momentum cutoff $g_{2/4}(q,\tau)\propto\Theta(q_\text{c}-q)$ the leading late-time behaviour will change to $E_\text{kin}(t)-E_\text{kin}^\infty\sim\sin[2\epsilon(q_\text{c})t]/t$ with a non-universal prefactor.\cite{RSM12}

\subsection{Fermionic Green function and quasiparticle weight}\label{sec:GF}
\subsubsection{Definition}
In the context of spinless fermions discussed in Sec.~\ref{sec:fermionicmodel} it is natural to consider the time evolution of the chiral Green function of the right movers
\begin{equation}
G_\text{F}(x,t)=\big\langle\Psi_+^\dagger(x,t)\,\Psi_+(0,t)\big\rangle,
\end{equation}
which has been studied in detail after sudden quenches in Refs.~\onlinecite{Cazalilla06,IucciCazalilla09,RSM12}. Using the representation of the right movers in terms of the bosonic modes \eqref{eq:RMtomodes} a straightforward calculation yields 
\begin{equation}
G_\text{F}(x,t)=\frac{\ii}{2\pi}\frac{1}{x+\ii 0}\exp\left(-\frac{1}{2}F_\text{F}(x,t)\right),
\label{eq:GFresult}
\end{equation}
where 
\begin{equation}
F_\text{F}(x,t)=4\int_0^\infty\frac{\dd q}{q}\bigl[1-\cos(qx)\bigr]\,\big|v(q,t)\big|^2
\label{eq:Fresult}
\end{equation}
encodes the deviation of the Green function from the non-interacting result. We note that the fermionic Green function \eqref{eq:GFresult} after a linear quench was already studied by Dora et al.~\cite{Dora-11} in a perturbative treatment in $g_2$ for $g_4=0$ (see also App.~\ref{app:PT}).

\subsubsection{Stationary limit}
First let us consider the stationary limit $F_\text{F}^\text{st}(x)=\lim_{t\to\infty}F_\text{F}(x,t)$, which reads 
\begin{equation}
F_\text{F}^\text{st}(x)=2 \int_0^\infty\frac{\dd q}{q}\bigl[1-\cos(qx)\bigr]\,\bigl(|A(q)|^2+|B(q)|^2\bigr).
\label{eq:Fst}
\end{equation}
We note in passing that the momentum dependence of the coupling functions and thus the coefficients $A(q)$ and $B(q)$ is essential for the convergence of the integral. The limiting behaviour at large distances $1/q_\text{c}\ll x$ is given by $F_\text{F}^\text{st}(x)=2\gamma_\text{F}\ln\big|q_\text{c}x\big|$ with the prefactor $\gamma_\text{F}$ taking the values
\begin{equation}
\gamma_\text{F}=\left\{\begin{array}{ll}
\displaystyle\gamma_\text{F}^\text{ad}=\frac{1}{2}\left(K+\frac{1}{K}-2\right),& x\ll 2v\tau,\\[3mm]
\displaystyle\gamma_\text{F}^\text{sq}=\frac{1}{4}\left(K^2+\frac{1}{K^2}-2\right),& 2v\tau\ll x.
\end{array}\right.
\label{eq:gammasq}
\end{equation}
To order $\mathcal{O}(\hat{g}_2^2)$ we observe $\gamma_\text{F}^\text{sq}=2\gamma_\text{F}^\text{ad}$ in agreement with the perturbative result.\cite{Dora-11}

\subsubsection{Light-cone effect}\label{sec:GFlightcone}
Calabrese and Cardy~\cite{CalabreseCardy06,CalabreseCardy07} first identified the light-cone or horizon effect after sudden quenches in conformal field theories. They also put forward a rather natural picture which is as follows: The quench creates quasiparticles in the system, which, if they originate from closely separated points, are quantum entangled. They then propagate semi-classically through the system with unique speed $v$. If the two quasiparticles in such an entangled pair arrive at the points $x_{1,2}$ at time $t$ they induce correlations, which in turn imply a sharp light cone in space time at $|x_1-x_2|=2vt$. Such light cones in correlation functions have been subsequently observed in numerical simulation on the Bose--Hubbard model\cite{LauchliKollath08,Barmettler-12,Carleo-14} as well as short- and long-ranged spin systems,\cite{Manmana-09,HaukeTagliacozzo13,Eisert-13,Bonnes-14} and experimentally in ultra-cold atomic gases\cite{Cheneau-12,Langen-13} and ions.\cite{Richerme-14,Jurcevic-14}

\begin{figure}[t]
    \centering
    \includegraphics[width=0.49\textwidth]{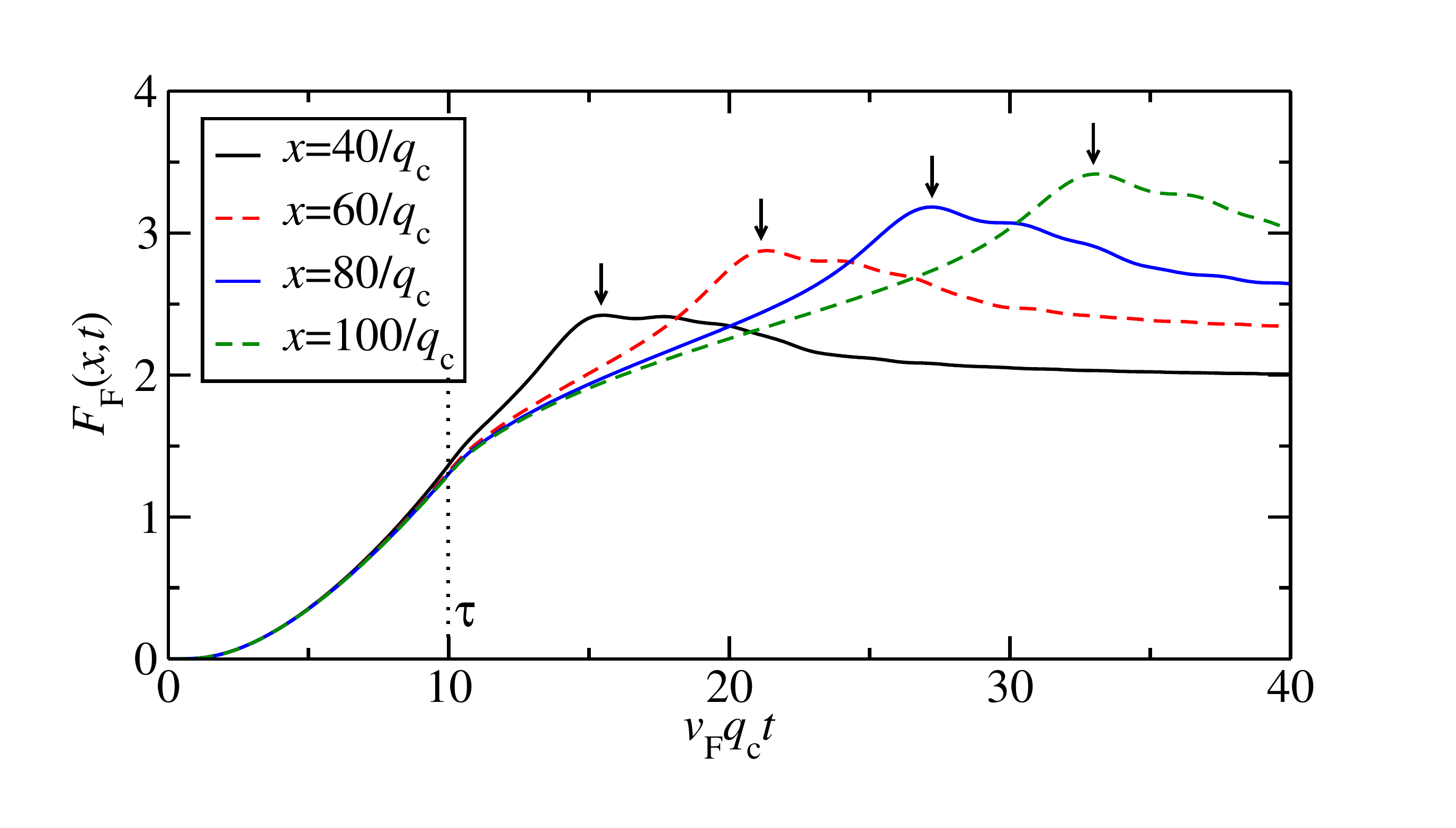}
    \caption{(Colour online) Cuts through the function \eqref{eq:Fresult} for fixed separations $x$ as a function of time $t$. We consider a linear quench with quench time $\tau=10/(v_\text{F}q_\text{c})$ and final interaction strength $g_0=2\pi v_\text{F}$. We observe a propagating maximum (indicated by arrows) which for sufficiently late times follows the linear relation $x=2\tilde{v}t$ with $\tilde{v}<v$.}
    \label{fig:GF}
\end{figure}
In order to analyse the light-cone effect in the fermionic Green function we first consider the time dependence at fixed separation $x$ as is often done to analyse numerical data as well;\cite{Manmana-09} exemplary results are shown in Fig.~\ref{fig:GF}. As can be seen there is a clear maximium (indicated by arrows) propagating through the system, which, for sufficiently late times, follows a linear relation of time vs position from which we can extract the velocity $\tilde{v}$ of the horizon via $\tilde{v}=x/(2t)$. The first observation is that this velocity is smaller than the renormalised velocity $v$ defined in \eqref{eq:LLparameter}, ie, $\tilde{v}<v$. The origin of this finding is the simple fact that for any non-trivial momentum dependence of the coupling functions the quasiparticles created by the quench will possess different group velocities $v(q)=\dd\epsilon(q)/\dd q$ depending on their individual momenta $q$. Since for monotonically falling coupling functions one has $v(q)<v$, the maximum originating from a propagating packet of quasiparticles will be delayed compared to the front of the fastest particles moving with $v$. This behaviour is already present for sudden quenches as we discuss in App.~\ref{app:1}. Since this delay of propagating quasiparticles is a general feature of any non-trivial momentum dependence of the coupling functions, it is also expected to show up in simulations for lattice models in both sudden and finite-time quenches (see also Fig.~\ref{fig:Ftilde}). 

\begin{figure}[t]
    \centering
    \includegraphics[width=0.49\textwidth]{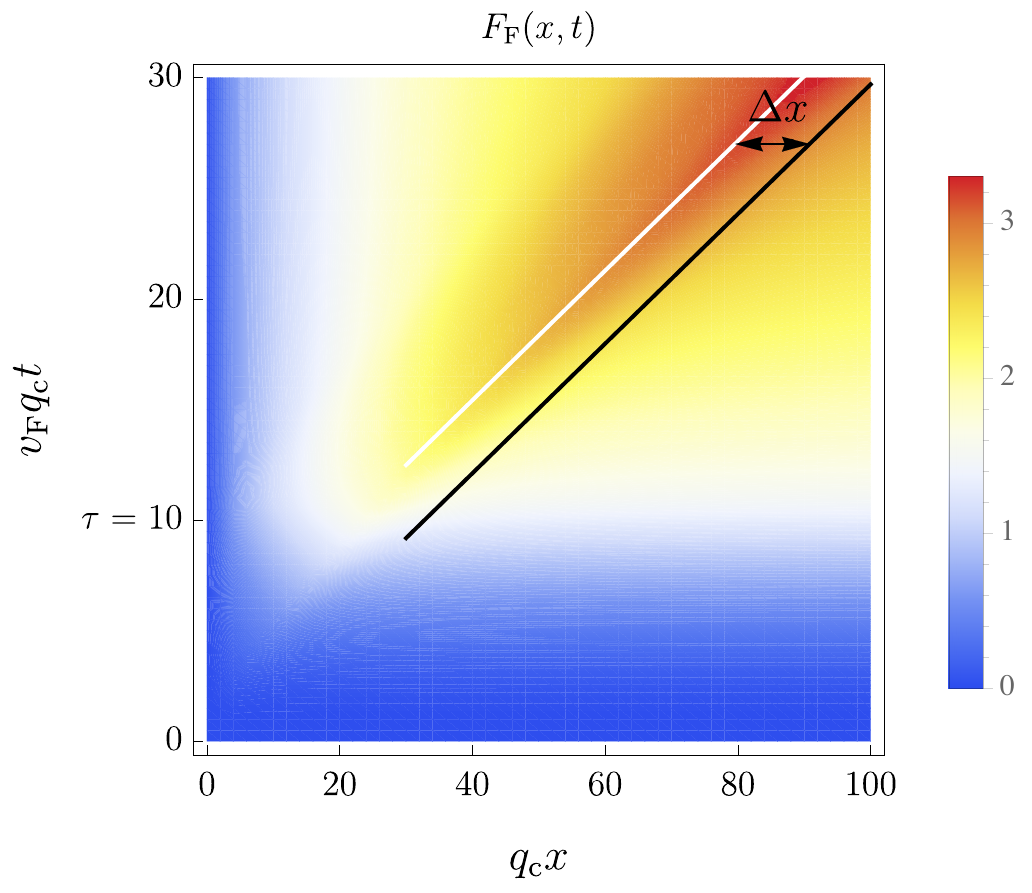}
    \caption{(Colour online) Contour plot of the function \eqref{eq:Fresult} after a linear quench of length $\tau=10/(v_\text{F}q_\text{c})$ and final interaction strength $g_0=2\pi v_\text{F}$. The white line indicates the light cone as identified by the propagating maximum shown in Fig.~\ref{fig:GF}, while the black line is the corresponding maximum after a sudden quench. We observe that after the quench the light cones propagate with identical velocities but that the maximum for the linear quench lags behind by a distance $\Delta x$. Here and in all following contour plots we used a linear interpolation between numerically evaluated data points.}
    \label{fig:GFcontour1}
\end{figure}
The space-time dependence of the function \eqref{eq:Fresult} after a linear quench is shown in Fig.~\ref{fig:GFcontour1}. The white line indicates the propagating maximum discussed above, while the black line is the corresponding maximum after a sudden quench with the same final parameters (see also App.~\ref{app:1}). At sufficiently late time after the quench both maxima propagate with identical velocities $\tilde{v}$. However, at any fixed time the maximum after the linear quench lags behind by a distance $\Delta x$. As we derive in App.~\ref{app:derivation}, the universal behaviour of \eqref{eq:Fresult} at late times and large separations, $\tau\ll t$ and $1/q_\text{c}, 2v\tau\ll x$, is given by
\begin{equation}
F_\text{F}(x,t)=F_\text{F}^\text{st}(x)-\gamma_\text{F}^\text{sq}\,\ln\left|1-\frac{x^2}{(2vt-\Delta x)^2}\right|,
\label{eq:Fapprox}
\end{equation}
where $F_\text{F}^\text{st}(x)$ is the stationary contribution defined in \eqref{eq:Fst}, the sudden-quench exponent $\gamma_\text{F}^\text{sq}$ was obtained in \eqref{eq:gammasq}, and the lag is given by
\begin{equation}
\Delta x=\frac{4K v_\text{F}\tau}{1-K^2}\left[\hat{g}_2(0,\tau)-\frac{1}{\tau}\int_0^\tau\dd t\,\hat{g}_2(0,t)\right].
\label{eq:lagapprox}
\end{equation}
These results are valid for short to moderate quench times, $v_\text{F}q_\text{c}\tau\lesssim 1$, but arbitrary quench protocols. For example, for linear or cosine quenches \eqref{eq:lagapprox} simplifies to $\Delta x=2Kv_\text{F}\hat{g}_2(0,\tau)\tau/(1-K^2)$. Furthermore, the approximate result \eqref{eq:Fapprox} only depends on the values $g_2(0,\tau)$ and $g_4(0,\tau)$; thus together with the stationary contribution $F_\text{F}^\text{st}(x)=2\gamma_\text{F}^\text{sq}\,\ln|q_\text{c}x|$ it describes the universal behaviour of the fermionic Green function at late times and large distances. For $\tau\to 0$ we recover the well-known result for sudden quenches.\cite{Cazalilla06} We note that since \eqref{eq:Fapprox} is obtained using the small-momentum expansion of $\epsilon(q)$ it  neglects the reduction of the propagation velocity from $v$ to $\tilde{v}$ (see App.~\ref{app:derivation} for a more detailed discussion).

As an alternative, heuristic ansatz to analyse the light-cone effect we use
\begin{equation}
\Delta x=2\tilde{v}\tau-2v_\text{F}\int_{s_\text{av}}^\tau\dd t'\,\sqrt{[1+\hat{g}_4(0,t')]^2-\hat{g}_2(0,t')^2}.
\label{eq:lag}
\end{equation}
Here the first term is due to the reduced post-quench evolution time during which the maximum propagates with velocity $\tilde{v}$. The second term describes the evolution during the quench, where we assume the quasiparticles to propagate with the instantaneous velocity~\cite{Bernier-14} $v_\text{F}\sqrt{[1+\hat{g}_4(0,t')]^2-\hat{g}_2(0,t')^2}$ (we have neglected the momentum dependence of the coupling functions for simplicity). Furthermore, the quasiparticles are created over the full quench time $0\le t\le\tau$, with the phenomenological parameter $s_\text{av}$ corresponding to the ``average" creation time of the relevant quasiparticles. Heuristically we find that $s_\text{av}$ grows with $\tau$ [$s_\text{av}\propto\tau$ for $v_\text{F}q_\text{c}\tau\lesssim 1$ in agreement with \eqref{eq:lagapprox}], decreases with increasing post-quench interaction strengths, and is larger for the cosine protocol than for the linear ramp. In principle, $s_\text{av}$ should be related to the mode occupations of the instantaneous eigenmodes $\alpha_n^t$ [where the parameter $t$ indicates that these modes diagonalise the Hamiltonian \eqref{eq:TLM} at time $t$ via $\alpha_n^t=c^t(q_n)b_n+s^t(q_n)b_{-n}^\dagger$], which is given by 
\begin{equation}
\bra{\Psi(t)}(\alpha_n^t)^\dagger\alpha_n^t\ket{\Psi(t)}=\big|c^t(q_n)v_n(t)+s^t(q_n)u_n(t)\big|^2
\label{eq:modes2}
\end{equation}
with the coefficients $s^t(q)$ and $c^t(q)$ given by \eqref{eq:defsc} with the coupling functions $g_{2/4}(q)$ taken at time $t$. However, the precise relation between \eqref{eq:modes2} and the parameter $s_\text{av}$ remains unclear.

\begin{figure}[t]
	\centering
	\includegraphics[width=0.49\textwidth]{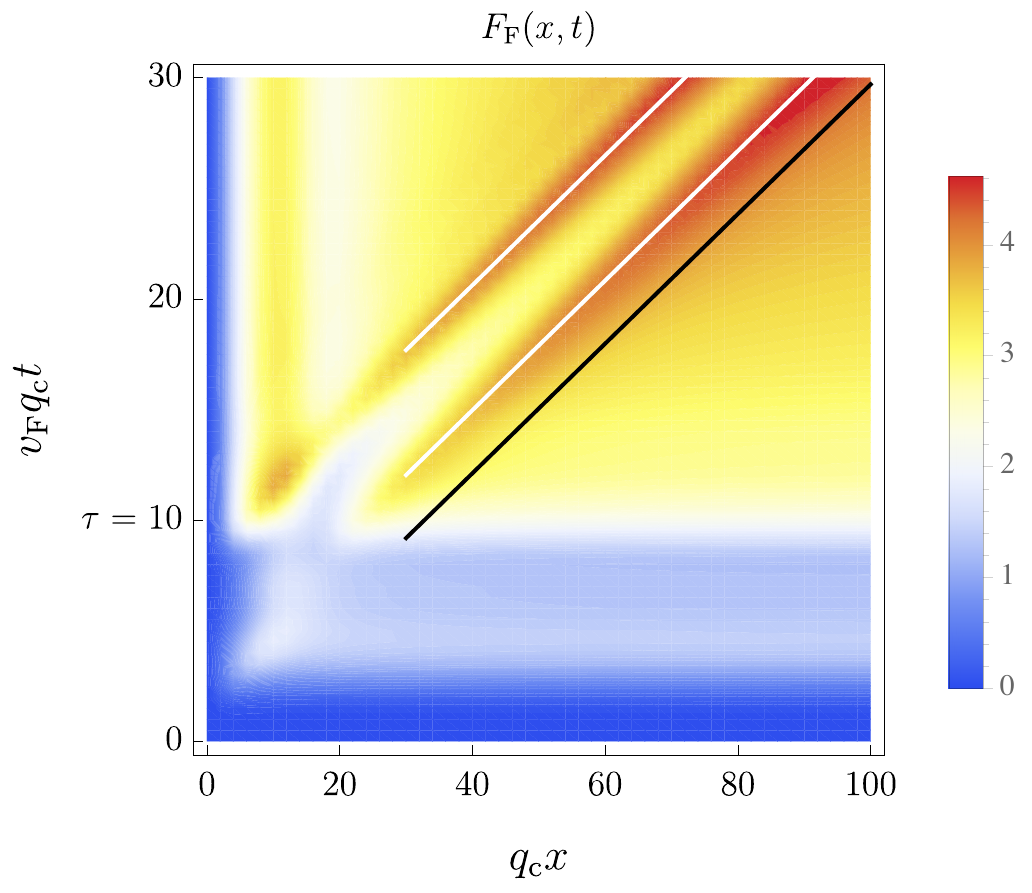}
	\caption{(Colour online) Contour plot of the function \eqref{eq:Fresult} after a periodic quench with $\nu=3$, length $\tau=10/(v_\text{F}q_\text{c})$ and $g_0=2\pi v_\text{F}$. The white lines indicate two light cones as identified by the propagating maxima, while the black line is the corresponding maximum after a sudden quench.}
	\label{fig:GFcontour2}
\end{figure}
Finally, in Fig.~\ref{fig:GFcontour2} we show the space-time dependence of the fermionic Green function after a periodic quench with long quench time $\tau=10/(v_\text{F}q_\text{c})$ [for which \eqref{eq:Fapprox} is not applicable]. We observe two propagating maxima caused by the non-trivial time dependence of the creation of quasiparticles during the quench. Also the propagating maxima are narrower than for the linear quench with the same quench time (shown in Fig.~\ref{fig:GFcontour1}) since the creation of quasiparticles happens during shorter time intervals.

\subsubsection{Oscillations inside the light cone}\label{sec:oscillations}
\begin{figure}[t]
	\centering
	\includegraphics[width=0.49\textwidth]{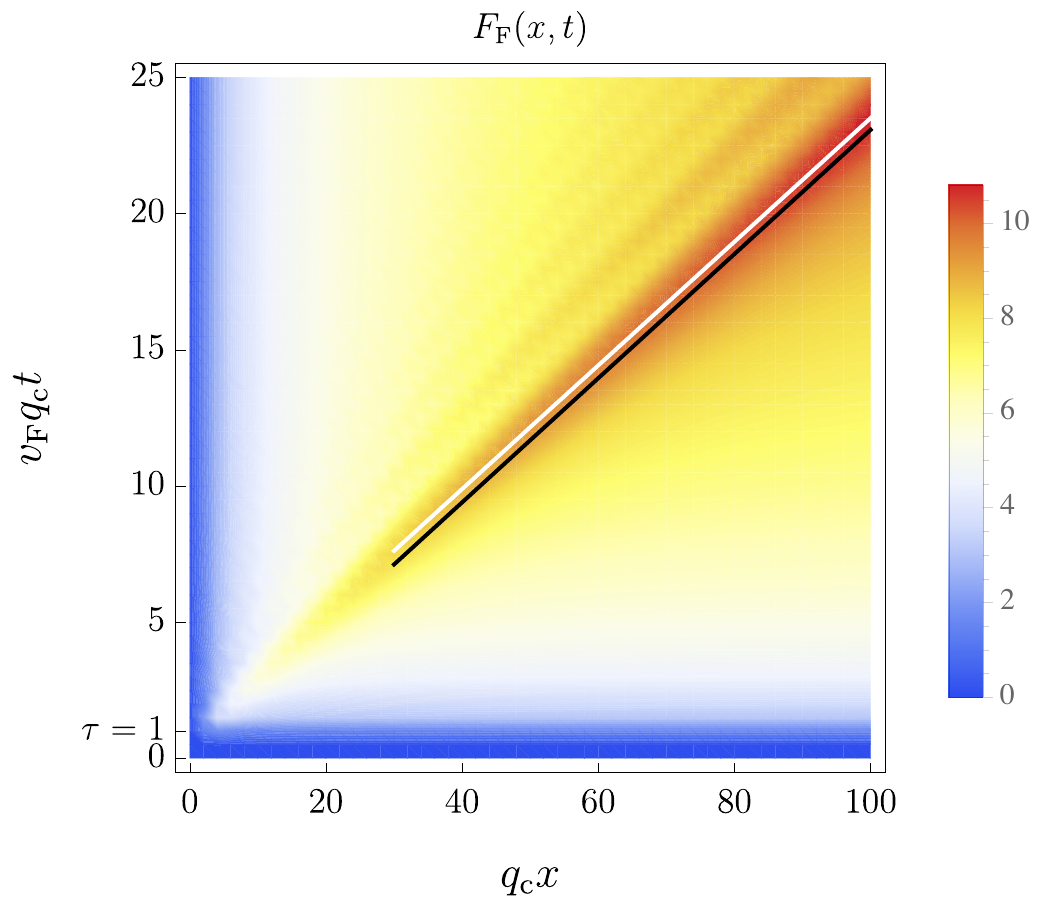}
	\caption{(Colour online) Contour plot of the function \eqref{eq:Fresult} after a cosine quench of length $\tau=1/(v_\text{F}q_\text{c})$ and $g_0=4\pi v_\text{F}$. In addition to the horizon we observe oscillations inside the light cone, which originate from the non-trivial momentum dependence of the coupling functions.}
	\label{fig:GFcontour3}
\end{figure}
The space-time dependence of the function \eqref{eq:Fresult} after a rather short cosine quench is shown in Fig.~\ref{fig:GFcontour3}. In addition to the horizon we observe oscillations inside the light cone, which become more pronounced when the quench rate $\hat{g}/\tau$ is increased. The origin of these oscillations is the non-trivial momentum dependence of the coupling functions and thus the single-mode energy $\epsilon(q)$, and hence eventually a result of the finite cutoff $q_\text{c}$ (see App.~\ref{app:oscillations} for more details). Since decreasing the quench time results in the creation of more quasiparticles at higher momenta (eg, compare the black and blue lines in Fig.~\ref{fig:modes}) where the momentum dependence of $\epsilon(q)$ is stronger, the oscillations are more pronounced for shorter quenches. We stress, however, that these oscillations are non-universal and indeed do not appear in the universal result \eqref{eq:Fapprox}. Nevertheless, the existence of a finite ultra-violet cutoff, eg, in lattice simulations, is generically expected to result in oscillating features following the horizon. For a discussion of similar, non-universal oscillations in the time evolution after sudden quenches we refer to Ref.~\onlinecite{RSM12} (see also Fig.~\ref{fig:GF-sudden}).

\subsubsection{Quasiparticle weight}\label{sec:Zfactor}
\begin{figure}[t]
    \centering
    \includegraphics[width=0.49\textwidth]{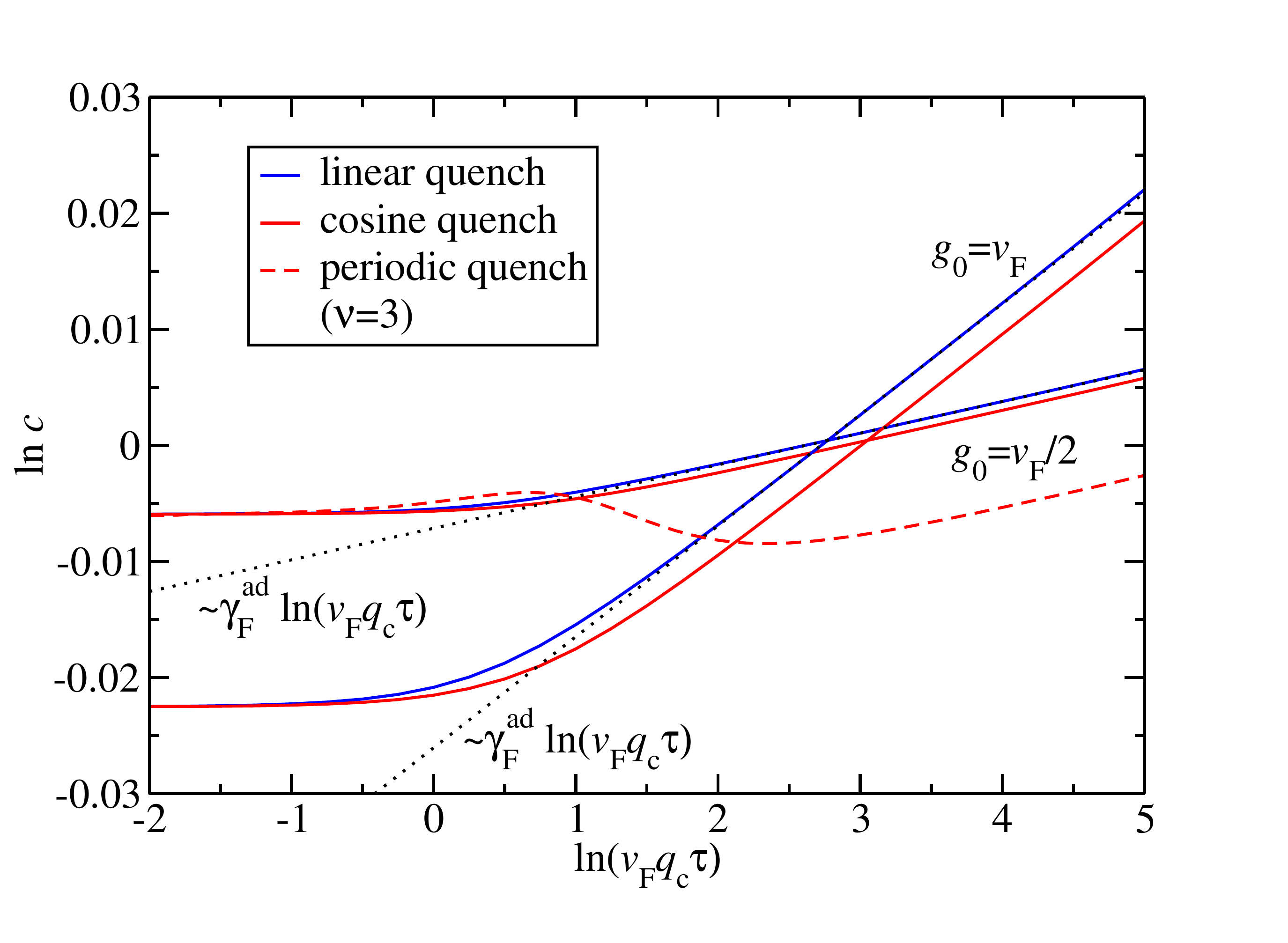}
    \caption{(Colour online) Quench-dependent prefactor $c$ in the quasiparticle weight \eqref{eq:Zfactor} after linear, cosine and periodic (only for $g_0=v_\text{F}/2$) quenches. For slow quenches, $v_\text{F}q_\text{c}\tau\gg 1$, we observe power-law enhancement $c\sim(v_\text{F}q_\text{c}\tau)^{\gamma_\text{F}^\text{ad}}$ indicated by the dotted lines. The occupation of high-energy modes after the periodic quench manifests itself in the non-monotonic behaviour around $v_\text{F}q_\text{c}\tau\sim 5$.}
    \label{fig:logZfactor}
\end{figure}

Finally we consider the fermionic quasiparticle weight $Z(t)$, whose late-time behaviour is characterised by universal power-law decay.\cite{Cazalilla06,IucciCazalilla09,Uhrig09,KRSM12,KRSM12,HamerlaUhrig13} In order to determine $Z(t)$ we consider the momentum distribution of right movers
\begin{equation}
n(q,t)=\int\dd x\,e^{\ii qx}\,G_\text{F}(x,t),
\end{equation}
which possesses a jump at the Fermi momentum $k_\text{F}$ with value $Z(t)=\lim_{q\to k_\text{F}-}n(q,t)-\lim_{q\to k_\text{F}+}n(q,t)$. At late times after the quench we find
\begin{equation}
Z(t)=c\,(v_\text{F}q_\text{c}\,t)^{-\gamma_\text{F}^\text{sq}},
\label{eq:Zfactor}
\end{equation}
in particular, the power-law decay is governed by the sudden-quench exponent $\gamma_\text{F}^\text{sq}$. However, the prefactor $c$ shows a dependence on the quench time $\tau$ as well as the quench protocol as shown in Fig.~\ref{fig:logZfactor}. For short and long quenches we further obtain the limiting behaviours $c\sim 1$ ($v_\text{F}q_\text{c}\tau\ll 1$) and $c\sim(v_\text{F}q_\text{c}\tau)^{\gamma_\text{F}^\text{ad}}$ ($v_\text{F}q_\text{c}\tau\gg 1$), in agreement with the perturbative results of Ref.~\onlinecite{Dora-11} for linear quenches.

\subsection{Bosonic Green function}\label{sec:SpSm}
\subsubsection{Definition}
Similarly to the fermionic Green function discussed above we consider its bosonic counterpart (see Sec.~\ref{sec:bosonicmodel})
\begin{equation}
G_\text{B}(x,t)=\big\langle\Psi_\text{B}(x,t)\,\Psi_\text{B}^\dagger(0,t)\big\rangle\propto\exp\left(-\frac{1}{2}F_\text{B}(x,t)\right)
\label{eq:GFB}
\end{equation}
where\cite{Pollmann-13,Bernier-14}
\begin{equation}
F_\text{B}(x,t)=\int_0^\infty\frac{\dd q}{q}\bigl[1-\cos(qx)\bigr]\,\big|u(q,t)-v(q,t)\big|^2.
\label{eq:Itt}
\end{equation}
In the limit of hard-core bosons the Green function \eqref{eq:GFB} corresponds to the spin-flip correlation function in the XXZ chain, which was studied by Pollmann et al.\cite{Pollmann-13} during linear quenches. More recently, Bernier et al.\cite{Bernier-14} analysed the bosonic Green function during a linear ramp of the interaction strength and identified the front at which correlations form [similar to the second term in Eq.~\eqref{eq:lag} above] as well as several regimes showing different power-law and stretched exponential decays. Here we will focus instead on the post-quench regime $t>\tau$. 

\subsubsection{Stationary limit}
First we consider the stationary limit $F_\text{B}^\text{st}(x)=\lim_{t\to\infty}F_\text{B}(x,t)$, which shows the  asymptotic behaviour
\begin{equation}
F_\text{B}^\text{st}(x)=2\gamma_\text{B}\,\ln(q_\text{c}x)
\label{eq:FBst}
\end{equation}
with the adiabatic and sudden-quench exponents
\begin{equation}
\gamma_\text{B}=\left\{\begin{array}{ll}
\displaystyle\gamma_\text{B}^\text{ad}=\frac{1}{2K},& x\ll 2v\tau,\\[3mm]
\displaystyle\gamma_\text{B}^\text{sq}=\frac{1}{4}\left(1+\frac{1}{K^2}\right),& 2v\tau\ll x.
\end{array}\right.
\label{eq:gammaBsq}
\end{equation}

\subsubsection{Light-cone effect and oscillations}
\begin{figure}[t]
  \includegraphics[width=0.49\textwidth]{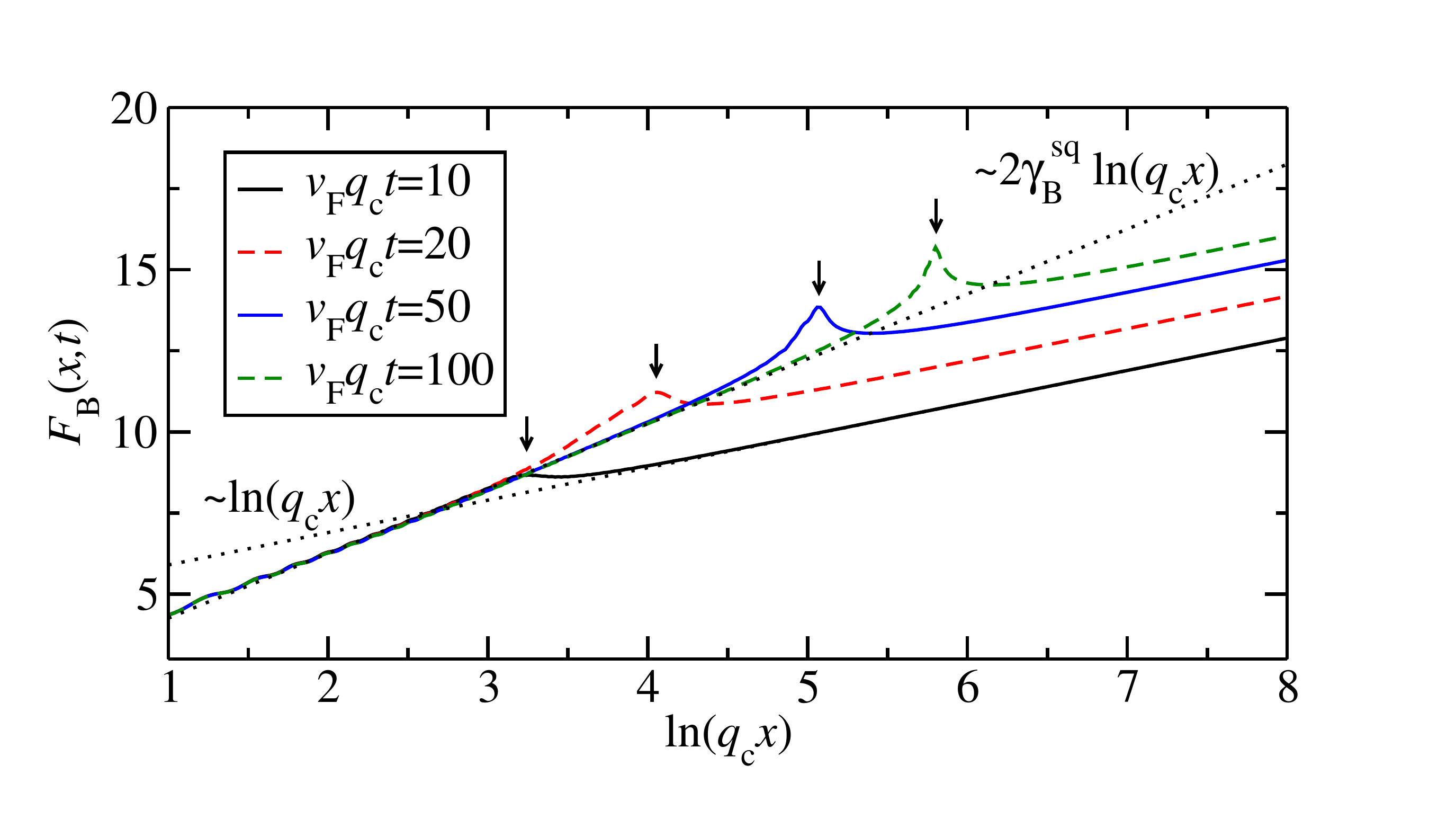}
  \caption{(Colour online) Constant time cuts for $F_\text{B}(x,t)$ after a linear quench of length $\tau=10/(v_\text{F}q_\text{c})$ and final interaction strength $g_0=2\pi v_\text{F}$. We observe a propagating maximum (indicated by arrows) defining the light cone. For large separations inside the light cone, $2v\tau\ll x\ll 2vt$, the slope is given by $2\gamma_\text{B}^\text{sq}$, while outside the light cone we find $F_\text{B}(x,t)\sim\ln(q_\text{c}x)$ implying the power-law decay of the bosonic Green function $G_\text{B}(x,t)\propto 1/\sqrt{x}$ in the non-interacting initial state.}
  \label{fig:GBcuts1}
\end{figure}
\begin{figure}[t]
  \includegraphics[width=0.49\textwidth]{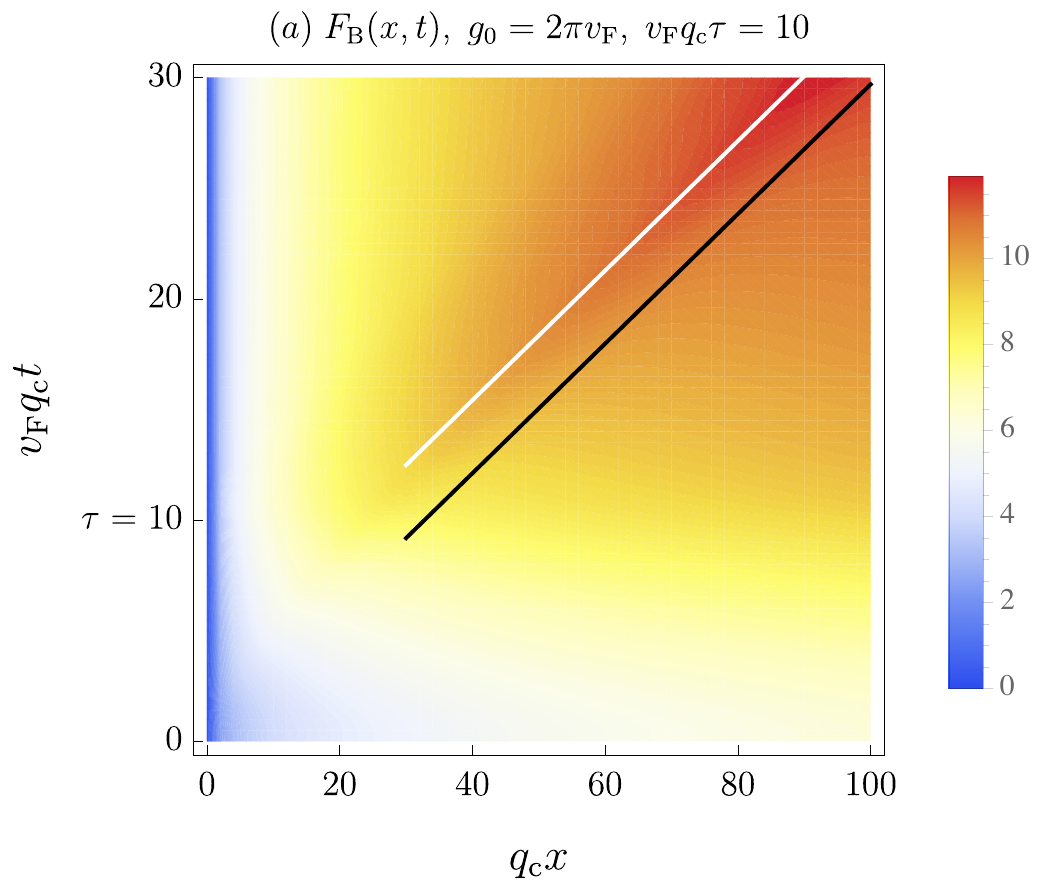}\vspace{5mm}
  \includegraphics[width=0.49\textwidth]{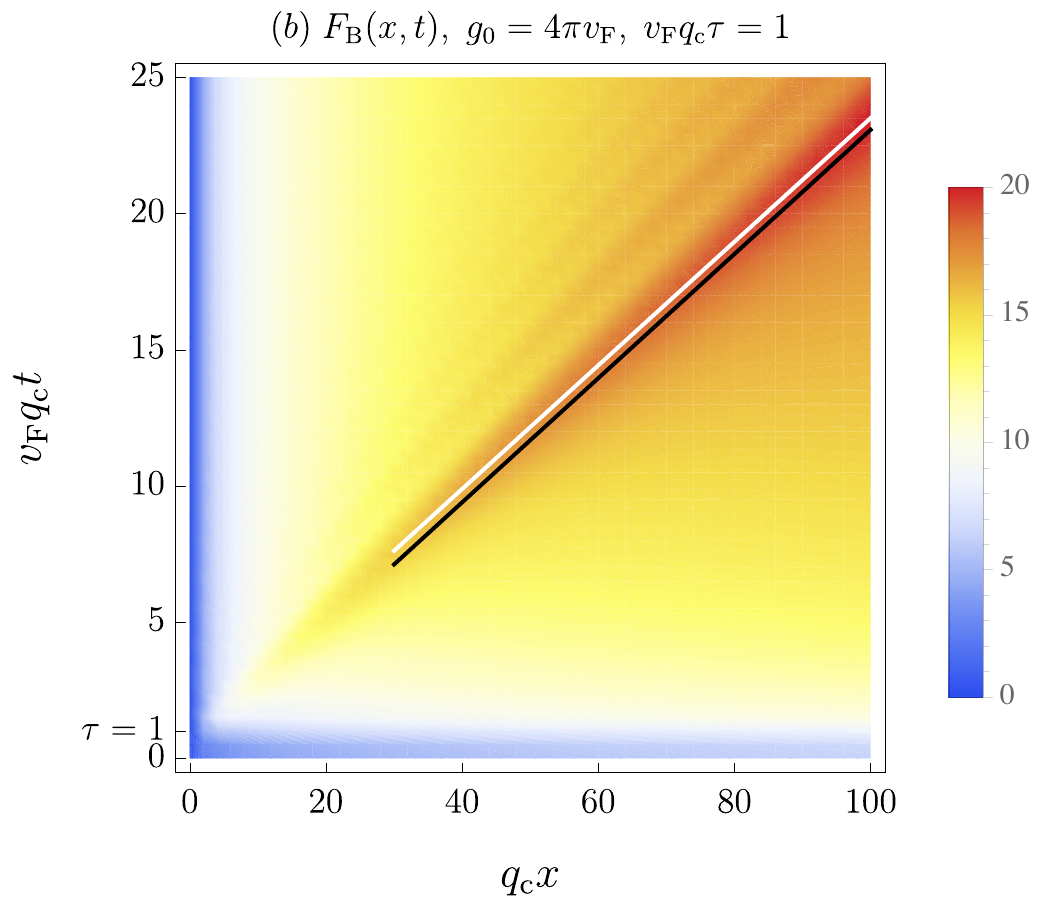}
  \caption{(Colour online) Contour plots of $F_\text{B}(x,t)$ after linear quenches with (a) $g_0=2\pi v_\text{F}$, $\tau=10/(v_\text{F}q_\text{c})$ and (b) $g_0=4\pi v_\text{F}$, $\tau=1/(v_\text{F}q_\text{c})$. The white lines indicate the light cones as identified by the propagating maximum shown in Fig.~\ref{fig:GBcuts1}, while the black lines are the corresponding maxima after a sudden quench. The feature for the linear quench lags behind by a distance $\Delta x$, which is identical to the one extracted from the fermionic Green function.}
  \label{fig:GBcontour1}
\end{figure}
The time evolution of $F_\text{B}(x,t)$ for linear quenches is shown in Figs.~\ref{fig:GBcuts1} and~\ref{fig:GBcontour1}. The propagating light cone is clearly visible in the cuts, the extracted position is identical to the one obtained from the fermionic Green function. In particular, we observe the same propagation velocity $\tilde{v}$ and the same lag $\Delta x$. The universal behaviour of $F_\text{B}(x,t)$ is obtained analogously to App.~\ref{app:derivation} with the result
\begin{equation}
F_\text{B}(x,t)=F_\text{B}^\text{st}(x)-\frac{1-K^2}{4K^2}\,\ln\left|1-\frac{x^2}{(2vt-\Delta x)^2}\right|,
\label{eq:FBapprox}
\end{equation}
with the lag given for short to moderate quench times but arbitrary quench protocols by \eqref{eq:lagapprox}. Together with \eqref{eq:FBst} we can obtain the space dependence of the bosonic Green function at fixed times (see also Fig.~\ref{fig:GBcuts1}): $G_\text{B}(x,t)\propto x^{-\gamma_\text{B}^\text{ad}}$ for $x\ll 2v\tau$, $G_\text{B}(x,t)\propto x^{-\gamma_\text{B}^\text{sq}}$ for $2v\tau\ll x\ll 2vt$, and $G_\text{B}(x,t)\propto x^{-1/2}$ for $2vt\ll x$, ie, outside the light cone. In addition, behind the propagating front we observe oscillations, see Fig.~\ref{fig:GBcontour1}(b), which, as for the fermionic Green function, originate from the non-trivial momentum dependence of the single-mode energy $\epsilon(q)$.

\subsubsection{Stretched exponential behaviour}\label{sec:strechedexponential}
Scrutinising a Galilean invariant system during linear ramps Bernier et al.~\cite{Bernier-14} identified an intermediate regime over which the bosonic Green function shows an unconventional stretched exponential space dependence at fixed times. 

In App.~\ref{app:stretchedexponential} we perform the similar analysis for the stationary Green function after a linear quench with $g_2(q,t)=g_4(q,t)=g_2(q)\,t/\tau$. We find that between the power-law dependencies in the adiabatic and sudden-quench regimes defined by \eqref{eq:gammaBsq}, there exists an intermediate regime showing the stretched exponential behaviour
\begin{equation}
G_\text{B}^\text{st}(x)\sim \exp\!\left[-\frac{2^{1/3}\pi^2\sqrt{1+2\hat{g}_2(0)}}{\Gamma(1/3)^3}\left(\frac{3\hat{g}_2(0)x}{v_\text{F}\tau}\right)^{1/3}\right]
\label{eq:GBstretched}
\end{equation}
provided
\begin{equation}
\frac{v_\text{F}\tau}{3\hat{g}_2(0)}\ll x\ll \frac{v_\text{F}\tau}{3\hat{g}_2(0)}\bigl[1+2\hat{g}_2(0)\bigr]^{3/2}.
\label{eq:stretchedregime}
\end{equation}
A few remarks are in order: (i) The existence of the regime \eqref{eq:stretchedregime} requires very strong post-quench interactions, which may not be realisable in microscopic models.\cite{Bernier-14} (ii) The result \eqref{eq:GBstretched} is only valid for linear quenches. (iii) In the derivation of \eqref{eq:GBstretched} we have used the replacement $g_2(q)\to g_2(0)$. However, given that the regime \eqref{eq:stretchedregime} corresponds to a regime of finite momenta and thus finite energies, the stretched exponential behaviour may be masked by the effects of marginal or irrelevant perturbations to the TLM like the momentum dependence of $g_2(q)$. (iv) Interestingly, the result \eqref{eq:GBstretched} is identical to the one\cite{Bernier-14} at $t=\tau$. Thus the emerging picture is as follows: During the quench the Green function develops the adiabatic and stretched exponential regimes (provided the post-quench interactions are strong enough) inside the light cone, while outside the light cone the behaviour is governed by the non-interacting initial state. After the quench the adiabatic and stretched exponential regimes remain unchanged, while behind the horizon the additional sudden-quench behaviour develops, eventually governing the whole regime $v_\text{F}\tau[1+2\hat{g}_2(0)]^{3/2}/[3\hat{g}_2(0)]\ll x$ in the stationary limit.

\subsection{Other correlation functions}
Using the same methods one can analyse the behaviour of other correlation functions. For example, the staggered part of the density-density correlation function is given by $\chi(x,t)\propto\exp\left(-\frac{1}{2}F_\chi(x,t)\right)$ with\cite{Pollmann-13} 
\begin{equation}
F_\chi(x,t)=\int_0^\infty\frac{\dd q}{q}\bigl[1-\cos(qx)\bigr]\,\big|u(q,t)+v(q,t)\big|^2.
\end{equation}
This shows the same qualitative features as the Green functions discussed in the previous two sections, ie, a clear light-cone effect with a delay due to the finite quench time as well as oscillations inside the light cone originating from the finite cutoff $q_\text{c}$. These general features are expected for other correlation functions as well.

\section{Conclusion and discussion}\label{sec:conclusion}
In this work we have investigated the time evolution in the TLM during and after finite-time interaction quenches. These were implemented by time dependent protocols to change the interaction parameters $g_{2/4}$ over the time interval $\tau$. After discussing the general framework of the time-dependent TLM, we derived exact analytical expressions for the small-momentum behaviour of the solution, as well as discussed the full solutions for specific quench protocols like the linear quench,\cite{Bernier-14} a cosine ramp and periodic driving.\cite{Pielawa11,BukovHeyl12}

We then used these results to analyse the time evolution of the total and kinetic energy as well as fermionic and bosonic Green functions during and after the quench. We focused on universal quantities in the sense that they only depend on the coupling functions at zero momentum and thus the Luttinger parameter and renormalised velocity given in Eq.~\eqref{eq:LLparameter}. For example, we showed that the kinetic energy decays as $\gamma_\text{kin}/t^{2}$ to its stationary value, where the decay parameter $\gamma_\text{kin}$ is identical to the one found for sudden quenches\cite{KRSM12} and thus independent of the quench protocol.

Analysing the stationary limit of the fermionic Green function we found a crossover from the adiabatic to the sudden regime at $x\sim 2v\tau$, where the two regimes are governed by different power-law decays, in agreement with earlier findings in the perturbative regime.\cite{Dora-11} Perhaps most interestingly, the light-cone effect\cite{CalabreseCardy06} well-known from sudden quenches is also clearly visible after finite-time quenches. However, as compared to the sudden case there is a lag of the horizon, which is related to two physical effects: First, during the quench the quasiparticles propagate at the instantaneous velocity\cite{Bernier-14} which is generically smaller than the post-quench velocity. Second, the creation of quasiparticle pairs happens during the full time of the quench, while for sudden quenches they are all created at the same time $t=0$. Using the analytical expressions for the small-momentum behaviour of the solution, we obtained the universal behaviour \eqref{eq:Fapprox} of the fermionic Green function. This includes an analytic expression for the lag \eqref{eq:lagapprox}, which is valid for short to moderate quench times, and relates the lag to the change of the interaction strength during the quench. In particular, the lag thus depends on the details of the quench protocol. Furthermore, we identified a reduction of the post-quench velocity with respect to the renormalised velocity as well as oscillations inside the light cone, and traced both effects back to the momentum dependence of the coupling functions $g_{2/4}(q)$.

Finally we analysed the bosonic Green function. The behaviour is very similar to the one discussed for the fermionic one. In particular, we extracted the universal behaviour of the post-quench dynamics, see \eqref{eq:FBst} and \eqref{eq:FBapprox}, and showed that the lag of the horizon is still given by \eqref{eq:lagapprox}. In addition, for linear quenches to very strong interactions we analysed the regime of intermediate separations. We found that the stretched exponential behaviour previously observed\cite{Bernier-14} during the quench is unaffected by the post-quench dynamics and thus also present in the stationary Green function.\pagebreak

As discussed in Sec.~\ref{sec:model} the TLM describes the universal low-energy physics of various one-dimensional fermionic and bosonic lattice models as well as spin chains in equilibrium, more specifically it corresponds to the low-energy fixed point in a renormalisation-group treatment. One may wonder to what extent results obtained for the quench dynamics in the TLM can also be used to analyse the quench dynamics in these lattice models, since the quench will inject a finite energy density into the system and thus drive it away from its low-energy fixed point. This may even be more severe in the case of finite-time quenches since the quench time $\tau$ will introduce an additional energy scale in the problem, which may increase the importance of marginal and irrelevant perturbations to the TLM. Nevertheless, various numerical studies\cite{DeChiara-06,Barmettler-09,Barmettler-10,KRSM12,KennesMeden13,HamerlaUhrig13,Coira-13,Collura-15} of observables in one-dimensional lattice models after sudden quenches showed a surprisingly good agreement with the results obtained in the TLM; a finding also obtained for the time evolution during finite-time interaction quenches\cite{Bernier-12,HaqueZimmer13,Bernier-14} in the Bose--Hubbard model. Still, from a practical point of view the study of the time evolution after finite-time quenches is complicated by the restriction of the achievable times in numerical simulations due to the finite quench time and the unknown effects of perturbations to the TLM as well as the energy (and thus time) scales involved. Further research in this direction is clearly desirable. 

In light of this it would be very interesting to investigate the effect of perturbations around the TLM, for example, the finite band curvature included in the non-linear Luttinger liquid theory\cite{Imambekov-12} or relevant perturbations leading to the opening of an excitation gap.\cite{Gritsev-07,IucciCazalilla10,BSE14} Furthermore, our results for periodic quenches could be used to connect the field of quantum quenches to periodically driven systems since the latter can be treated by increasing the number of periods $\nu$ in \eqref{eq:cosine3quench} as already employed in Refs.~\onlinecite{Pielawa11,BukovHeyl12}.

\acknowledgements
We thank Jean-S\'{e}bastien Bernier, Bal\'{a}zs D\'{o}ra, Masud Haque, Markus Heyl, Salvatore Manmana, Volker Meden and Tatjana Pu\v{s}karov for useful discussions, and Nicholas Ohs for collaboration in the very early stages of this project. This work is part of the D-ITP consortium, a program of the Netherlands Organisation for Scientific Research (NWO) that is funded by the Dutch Ministry of Education, Culture and Science (OCW). This work was  supported by the German Research Foundation (DFG) through the Emmy-Noether Program under SCHU 2333/2-1, and the Foundation for Fundamental Research on Matter (FOM), which is part of the Netherlands Organisation for Scientific Research (NWO), under 14PR3168.

\appendix
\section{Solution at small momenta}\label{app:smallk}
In this appendix we obtain the solution of \eqref{eq:DGLuv} at small momenta. To this end we expand the solutions in powers of $q/q_\text{c}$, $u(q,t)=\sum_{m=0}^\infty u^{(m)}(t)\,(q/q_\text{c})^m$ and $v(q,t)=\sum_{m=0}^\infty v^{(m)}(t)\,(q/q_\text{c})^m$. Doing the same for the coefficients $\omega(q,t)$ and $\lambda(q,t)$ [recall that, eg, $\omega_n(t)=\omega(q_n,t)$] we obtain the equations $\dot{u}^{(0)}(t)=\dot{v}^{(0)}=0$, $\ii\dot{u}^{(1)}(t)=v_\text{F}q_\text{c}[1+\hat{g}_4(0,t)]$ and $\ii\dot{v}^{(1)}(t)=-v_\text{F}q_\text{c}\hat{g}_2(0,t)$ with the initial conditions $u^{(0)}(0)=1$, $v^{(0)}(0)=u^{(1)}(0)=v^{(1)}(0)=0$. This immediately results in \eqref{eq:smallku} and \eqref{eq:smallkv}.

In order to compare the leading and next-to-leading term in the expansion \eqref{eq:smallku}, we approximate the integral by taking $\hat{g}_4(0,t')\to\hat{g}_4(0,\tau)=\hat{g}_4(0)$, which results in $u(q,\tau)=1-\ii v_\text{F}q\tau[1+\hat{g}_4(0)]$. Thus we see that the first and second term become of the same order for $q\sim 1/(v_\text{F}\tau)/[1+\hat{g}_4(0)]\approx 1/(v_\text{F}\tau)$, thus establishing the requirement $q\ll 1/(v_\text{F}\tau)$. Of course, considering very non-monotonic time dependencies of $\hat{g}_4(0,t)$ may lead to a shrinking of the range of momenta over which \eqref{eq:smallku} is applicable. 

It is straightforward to determine the behaviour in second order. We get $u^{(2)}(t)=(|v^{(1)}(t)|^2-|u^{(1)}(t)|^2)/2$, while the next-to-leading contribution to $v(q,t)$ is given by
\begin{equation}
\begin{split}
v^{(2)}(t)=&(v_\text{F}q_\text{c})^2\int_0^t\dd t'\int_0^{t'}\dd t''\,\Bigl[\hat{g}_2(0,t')\bigl(1+\hat{g}_4(0,t'')\bigr)\\
&\qquad\qquad\qquad-\bigl(1+\hat{g}_4(0,t')\bigr)\hat{g}_2(0,t'')\Bigr]\\
&+\ii v_\text{F}q_\text{c}^2\int_0^t\dd t'\,\frac{\partial}{\partial q}\hat{g}_2(q,t')\Big|_{q=0}.
\end{split}
\label{eq:v2}
\end{equation}
To obtain an estimate for \eqref{eq:v2} we assume a linear quench with $g_2(q,t)=g_4(q,t)=g_2(q)\,t/\tau$ and $\partial g_2(q)/\partial q|_{q=0}=0$. This yields 
\begin{equation}
v(q,\tau)=\frac{\ii v_\text{F}\hat{g}_2(0)\tau q}{2}+\frac{v_\text{F}^2\hat{g}_2(0)\tau^2 q^2}{6}+\ldots 
\end{equation}
[We stress that the second-order term is still linear in $\hat{g}_2(0)$, ie, the expansion is non-perturbative.] Thus we see that the first and second-order terms become comparable for $q\sim 1/(v_\text{F}\tau)$, thus again  leading to the requirement $q\ll 1/(v_\text{F}\tau)$. The same estimate is found for other quench protocols. We note again that considering very non-monotonic time dependencies of the quench protocol or very strong final interaction strengths may lead to a shrinking of the range of momenta over which \eqref{eq:smallku} and \eqref{eq:smallkv} is applicable. 

\section{Perturbative solution}\label{app:PT}
For completeness we state here the perturbative solution of \eqref{eq:DGLuv}, which up to second order in $\hat{g}_{2/4}(q_n,t)$ reads
\begin{eqnarray}
u_n(t)&=&e^{-\ii v_\text{F}|q_n|t}-\ii v_\text{F}|q_n|\int_0^t\dd t'\,\hat{g}_4(q_n,t')\,e^{-\ii v_\text{F}|q_n|t}\nonumber\\*
& &\hspace{-5mm}
-(v_\text{F}q)^2e^{-\ii v_\text{F}|q|t}\int_0^t\dd t'\int_0^{t'}\dd t''\,\Bigl[\hat{g}_4(q,t')\,\hat{g}_4(q,t'')\nonumber\\*
& &\qquad-\hat{g}_2(q,t')\,\hat{g}_2(q,t'')e^{2\ii v_\text{F}|q|(t'-t'')}\Bigr],\\
v_n(t)&=&\ii v_\text{F}|q_n|\int_0^t\dd t'\,\hat{g}_2(q_n,t')\,e^{\ii v_\text{F}|q_n|(t-2t')}\nonumber\\*
& &\hspace{-5mm}
+(v_\text{F}q)^2e^{\ii v_\text{F}|q|t}\int_0^t\dd t'\int_0^{t'}\dd t''\,
\Bigl[\hat{g}_2(q,t')\,\hat{g}_4(q,t'')\quad\nonumber\\*
& &\times e^{-2\ii v_\text{F}|q|t'}-\hat{g}_4(q,t')\,\hat{g}_2(q,t'')e^{-2\ii v_\text{F}|q|t''}\Bigr].\quad
\label{eq:PTsolution}
\end{eqnarray}
For $g_4=0$ and restricting to $\mathcal{O}(\hat{g}_2)$ we recover the result given in Ref.~\onlinecite{Dora-11}.

\section{Light-cone velocity after sudden quenches}\label{app:1}
\begin{figure}[t]
	\centering
	\includegraphics[width=0.49\textwidth]{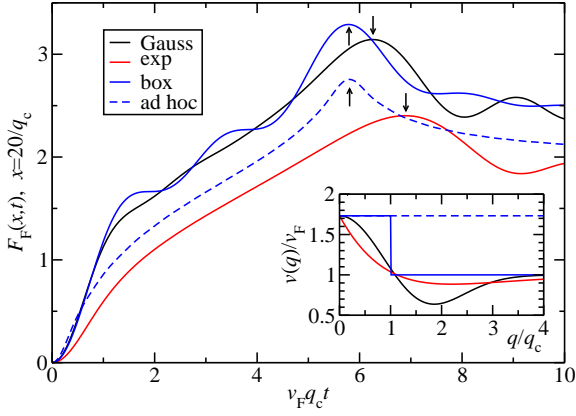}
	\caption{(Colour online) Time evolution of the function \eqref{eq:Fresult} at $x=20/q_\text{c}$ after a sudden quench from $g_0=0$ to $g_0=2\pi v_\text{F}$. We assumed $g_2(q,\tau)=g_4(q,\tau)$ and considered Gaussian momentum dependence $g_2(q,\tau)=g_0\,\exp[-(q/q_\text{c})^2/2]$, exponential dependence $g_2(q,\tau)=g_0\,\exp[-q/q_\text{c}]$, a box potential $g_2(q,\tau)=g_0\,\Theta(q_\text{c}-q)$, or the ad-hoc regularisation (see text). The arrows indicate the positions of the propagating maxima. For the Gaussian and exponential coupling functions we observe a delay of the maximum caused by the reduced group velocities $v(q)=\dd\epsilon(q)/\dd q$ of the quasiparticles (shown in the inset). The result for the box potential shows clear oscillations inside the light cone originating from the finite, sharp cutoff $q_\text{c}$.}
	\label{fig:GF-sudden}
\end{figure}
In this appendix we briefly discuss the propagation of the horizon in the fermionic Green function after a sudden quench. In Fig.~\ref{fig:GF-sudden} we compare the time dependence at fixed separation for different momentum dependencies of the coupling functions, including the ad-hoc regularisation~\cite{Cazalilla06,IucciCazalilla09} which allows an analytic treatment of the momentum integrals. The latter is defined by choosing the coupling functions such that $4s(q)^2c(q)^2=\hat{g}_2(0)^2 e^{-q/q_\text{c}}/W(0)^2$, and in addition linearising the dispersion relation $\epsilon(q)\to v|q|$ (see Ref.~\onlinecite{RSM12} for more details). Thus in the ad-hoc regularisation all quasiparticles travel with the renormalised velocity $v$ independently of their individual momenta. The same is true for the relevant momenta $|q|<q_\text{c}$ in the case of the box potential. In contrast, if the coupling functions possess a genuine momentum dependence the effective velocities $v(q)=\dd\epsilon(q)/\dd q$ of individual quasiparticles are reduced, as is shown in the inset to Fig.~\ref{fig:GF-sudden}. Hence we expect a delay of propagating features for momentum-dependent coupling functions as exemplified in the main panel of Fig.~\ref{fig:GF-sudden} for the propagating maxima indicated by the arrows. This effect is also expected to show up in lattice systems like spin chains, where the ultra-violet cutoff provided by the lattice causes an effective momentum dependence of the effective coupling functions.

\section{Derivation of Eq.~\ref{eq:Fapprox}}\label{app:derivation}
In this appendix we extract the universal behaviour of the function \eqref{eq:Fresult} at late times and large distances. Since $t>\tau$ we first insert the solution \eqref{eq:vafterquench} after the quench to obtain
\begin{eqnarray}
F_\text{F}(x,t)&=&F_\text{F}^\text{st}(x)+\tilde{F}_\text{F}(x,t),\\
\tilde{F}_\text{F}(x,t)&=&2\int_0^\infty\frac{\dd q}{q}\bigl[1-\cos(qx)\bigr]\nonumber\\*
& &\times\Bigl[\bigl(|A(q)|^2-|B(q)|^2\bigr)\cos[2\epsilon(q)t]\nonumber\\*
& &\quad+2\,\text{Re}\bigl[A(q)^*B(q)\bigr]\,\sin[2\epsilon(q)t]\Bigr],\label{eq:appFtilde}
\end{eqnarray}
where $F_\text{F}^\text{st}(x)=\lim_{t\to\infty}F_\text{F}(x,t)$ denotes the stationary limit given by \eqref{eq:Fst}. The integral in \eqref{eq:appFtilde} is dominated by small momenta $q\ll q_\text{c}$, for which we can use the expansions \eqref{eq:smallkA} and \eqref{eq:smallkB}. The additional requirement $q\ll 1/(v_\text{F}\tau)$ originating from \eqref{eq:smallku} and \eqref{eq:smallkv} will be automatically satisfied provided $v_\text{F}q_\text{c}\tau\lesssim 1$, implying that the results are valid for quenches with short to moderate quench times. Writing  $A(q)=-\ii A_1 q$ and $B(q)=\ii B_0$ with $A_1,B_0\in\mathbb{R}$ we obtain
\begin{eqnarray}
\tilde{F}_\text{F}(x,t)&=&2\int_0^\infty\frac{\dd q}{q}\bigl[\cos(qx)-1\bigr]\\*
& &\times\Bigl[B_0^2\cos[2\epsilon(q)t]+2A_1B_0q\,\sin[2\epsilon(q)t]\Bigr]\nonumber\\
&&\hspace{-15mm}
=2B_0^2\int_0^\infty\frac{\dd q}{q}\bigl[\cos(qx)-1\bigr]\cos\!\!\left[2\epsilon(q)t-\frac{2A_1}{B_0} q\right]\!\!.\quad\label{eq:Ftildeapp4}
\end{eqnarray}
Finally, expanding the single-mode energy to leading order, $\epsilon(q)=vq+\ldots$, gives
\begin{equation}
\tilde{F}_\text{F}(x,t)=2B_0^2\int_0^\infty\frac{\dd q}{q}\bigl[\cos(qx)-1\bigr]\cos\bigl[(2vt-\Delta x)q\bigr]
\end{equation}
with $\Delta x=2A_1/B_0$. The remaining integral can be performed analytically\cite{AbramowitzStegun65} with the result given in Eq.~\eqref{eq:Fapprox}. We note that the approximation $\epsilon(q)=vq$ neglects all effects originating from the momentum dependence of the group velocity $v(q)=\dd\epsilon(q)/\dd q$ such as oscillations inside the light cone (see App.~\ref{app:oscillations}) or corrections to the sudden-quench exponent $\gamma_\text{F}^\text{sq}=B_0^2$ (similar to the corrections discussed by Meden\cite{Meden99} for the one-particle Green function in equilibrium).

\section{Oscillations inside the light cone}\label{app:oscillations}
\begin{figure}[t]
	\centering
	\includegraphics[width=0.49\textwidth]{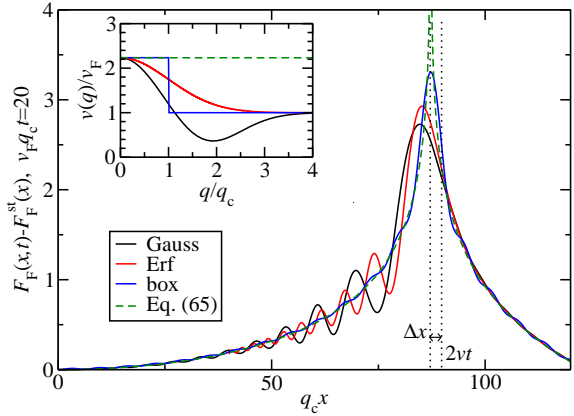}
	\caption{(Colour online) Time evolution of the function \eqref{eq:Fresult} at $t=20/(v_\text{F}q_\text{c})$ after a linear quench of length $\tau=1/(v_\text{F}q_\text{c})$ to $g_0=4\pi v_\text{F}$. The momentum dependencies of the coupling function $g_2(q,\tau)=g_4(q,\tau)$ are described in the text. In addition, we show the universal result \eqref{eq:Fapprox}. The dotted lines indicate the position of the horizon after a sudden quench ($x=2vt$) as well as in the universal result, where it is given by \eqref{eq:lagapprox}. Inset: Effective quasiparticle velocities $v(q)=\dd\epsilon(q)/\dd q$.}
	\label{fig:Ftilde}
\end{figure}
In Fig.~\ref{fig:Ftilde} we show constant-time cuts of the function \eqref{eq:Fresult} after a linear quench. We compare the results for two specific momentum dependencies of the coupling functions, namely the Gaussian momentum dependence $g_2(q,\tau)=g_0\,\exp[-(q/q_\text{c})^2/2]$ and one containing the error function\cite{AbramowitzStegun65} (denoted by Erf), ie, $g_2(q,\tau)=2\pi v_\text{F}\,\tilde{g}(q)\,[1+\tilde{g}(q)/2]$ with the auxiliary function
\begin{equation}
\tilde{g}(q)=\sqrt{\frac{\pi}{2}}\left(\sqrt{1+\frac{g_0}{\pi v_\text{F}}}-1\right)\frac{q_\text{c}}{q}\,\text{Erf}\left(\frac{q}{\sqrt{2}q_\text{c}}\right).
\label{eq:tildeg}
\end{equation}
The momentum dependence in \eqref{eq:tildeg} is chosen such that the group velocity is a monotonically decreasing function of $q$, ie, $v(q)=(v-v_\text{F})\,\exp[-(q/q_\text{c})^2/2]+v_\text{F}$. We also show result for the box potential $g_2(q,\tau)=g_0\,\Theta(q_\text{c}-q)$ as well as the universal result \eqref{eq:Fapprox}.

We observe that there are two main effects of the non-trivial momentum dependence: First, the reduced quasiparticle velocities $v(q)$ (see inset) lead to a reduction of the effective velocity of the horizon from $v$ to $\tilde{v}$, as can be clearly seen for the Gaussian and Erf momentum dependencies. Second, for these two cases we also observe pronounced oscillations inside the light cone. In the case of the box potential the finite cutoff also results in weak features inside the light cone, which are, however, much less pronounced. In contrast, the universal result \eqref{eq:Fapprox} does not show these two effects. However, oscillations show up if also the next-to-leading term in the expansion of the single-mode energy $\epsilon(q)$ in \eqref{eq:Ftildeapp4} is kept. From this we deduce that the appearance of oscillations is caused by the momentum dependence of the group velocity $v(q)$, for example by a finite curvature at $q=0$. Of course, the precise details of the oscillations, like the oscillation frequency and their decay as a function of $2\tilde{v}t-x$, depend on the full momentum dependence of the velocity as well as the prefactors $A(q)$ and $B(q)$ (and thus on the details of the quench).

\section{Stretched exponential behaviour}\label{app:stretchedexponential}
In this appendix we consider the time evolution of the bosonic Green function after linear quenches in the Galilean invariant system, ie, we have $g_2(q,t)=g_4(q,t)=g_2(q)\,t/\tau$ during the quench and $g_2(q,\tau)=g_4(q,\tau)=g_2(q)$ afterwards. Since $g_2(q,t)=g_4(q,t)$, we can rewrite the function \eqref{eq:Itt} in terms of the auxiliary function $a(q,t)$ using \eqref{eq:relationua} and \eqref{eq:relationva} as\cite{Bernier-14}
\begin{equation}
F_\text{B}(x,t)=\frac{1}{v_\text{F}^2}\int_0^\infty\frac{\dd q}{q^3}\bigl[1-\cos(qx)\bigr]\,\left|\frac{\dd}{\dd t}a(q,t)\right|^2.
\end{equation}
The solution of \eqref{eq:DGLa} after the quench is given by
\begin{equation}
a(q,t)=a(q,\tau)\cos\bigl[\epsilon(q)(t-\tau)\bigr]+\frac{\dot{a}(q,\tau)}{\epsilon(q)}\sin\bigl[\epsilon(q)(t-\tau)\bigr],
\end{equation}
which yields
\begin{equation}
\begin{split}
&\frac{\dd}{\dd t}a(q,t)=\frac{1}{2}\Bigl(\big|\dot{a}(q,\tau)\big|^2+\epsilon(q)^2\,\big|a(q,\tau)\big|^2\Bigr)\\*
&\quad+\frac{1}{2}\Bigl(\big|\dot{a}(q,\tau)\big|^2-\epsilon(q)^2\,\big|a(q,\tau)\big|^2\Bigr)\cos\bigl[2\epsilon(q)(t-\tau)\bigr]\\*
&\quad-\epsilon(q)\,\text{Re}\bigl[\dot{a}(q,\tau)^*\,a(q,\tau)\bigr]\sin\bigl[2\epsilon(q)(t-\tau)\bigr].
\end{split}
\end{equation}
Thus we can write the stationary part as
\begin{equation}
\begin{split}
F_\text{B}^\text{st}(x,t)=&\frac{1}{2v_\text{F}^2}\int_0^\infty\frac{\dd q}{q^3}\bigl[1-\cos(qx)\bigr]\\*
&\qquad\times\Bigl(\big|\dot{a}(q,\tau)\big|^2+\epsilon(q)^2\,\big|a(q,\tau)\big|^2\Bigr).
\end{split}
\label{eq:appFBst}
\end{equation}
The remaining analysis is now based on the exactly known functions $a(q,\tau)$ and $\dot{a}(q,\tau)$ after a linear quench, and the asymptotic expansions of the Bessel functions. It follows the derivation by Bernier et al.\cite{Bernier-14} one-to-one, thus we refrain from repeating it here. 


\end{document}